\documentclass{aa}  

\usepackage{graphicx}
\usepackage{amsmath}
\usepackage{txfonts}
\usepackage{natbib}
\usepackage{subfig}
\usepackage[T1]{fontenc}
\usepackage{times}
\bibpunct{(}{)}{;}{a}{}{,}

\newcommand{\beq} {\begin{equation}}
\newcommand{\eeq}{\end{equation}}
\newcommand{\bea} {\begin{eqnarray}}
\newcommand{\eea}{\end{eqnarray}}

\begin{document} 

   \title{Cosmic ray propagation in sub-Alfv\'enic magnetohydrodynamic turbulence}

   \author{R. Cohet
          \inst{1}
          \and
          A. Marcowith \inst{1}
          }

   \institute{Laboratoire Univers et particules de Montpellier,\
   Universit\'e de Montpellier, CNRS/IN2P3, CC 72, Place Eug\`ene Bataillon, \\
   34095 Montpellier Cedex 5, France\\ 
              \email{Alexandre.Marcowith@umontpellier.fr}
              }

   \date{Received -; accepted -}

 
  \abstract{
   {Context.} The propagation of cosmic rays or energetic charged particles in magnetized turbulence is a complex problem which involves non-linear wave-particle interactions and chaotic magnetic field lines transport. This problem has been addressed until recently using either analytical calculations or simulations using prescribed turbulence models. With the advent of super computers it is now possible to investigate energetic charged particle propagation using direct computation of electromagnetic fields. This is in particular the case for high-energy particles propagation in magnetohydrodynamic turbulence. \\ 
   {Aims.} This work has the main objective to provide a detailed investigation of cosmic ray propagation in magnetohydrodynamic turbulent fields generated by forcing the fluid velocity field at large scales. It provides a derivation of the particle mean free path dependences in terms of the turbulence level described by the Alfv\'enic Mach number and in terms of the particle rigidity. \\
   {Methods.} We use an upgrade version of the magnetohydrodynamic code {\tt RAMSES} which includes a forcing module and a kinetic module and solve the Lorentz equation for each particle. The simulations are performed using a 3 dimension periodical box in the test-particle and magnetostatic limits. The forcing module is implemented using an Ornstein-Uhlenbeck process. An ensemble average over a large number of particle trajectories is applied to reconstruct the particle mean free paths. \\
   {Results.} We derive the cosmic ray mean free paths in terms of the Alfv\'enic Mach numbers and particle reduced rigidities in different turbulence forcing geometries. The reduced particle rigidity is $\rho=r_L/L$ where $r_L$ is the particle Larmor radius and $L$ is the simulation box length related to the turbulence coherence or injection scale $L_{inj}$ by $L \sim 5 L_{inj}$.  We have investigated with a special attention compressible and solenoidal forcing geometries. \\
   {Conclusions.} We find that compressible forcing solutions are compatible with the quasi-linear theory or more advanced non-linear theories which predict a rigidity dependence as $\rho^{1/2}$ or $\rho^{1/3}$.  Solenoidal forcing solutions at least at low or moderate Alfv\'enic numbers are not compatible with the above theoretical expectations and require more refined arguments to be interpreted. It appears especially for Alfv\'enic Mach numbers close to one that the wandering of field lines controls perpendicular mean free path solutions whatever the forcing geometry. 
}
 \keywords{Physical data and processes: Magnetohydrodynamics (MHD)-Turbulence--Methods: numerical--Interstellar medium: Cosmic Rays}

   \maketitle
%

\section{Introduction}
\label{S:INT}
Cosmic rays (CRs) propagate in our Galaxy through the interaction with electromagnetic perturbations usually described in the magnetohydrodynamic (MHD) approximation \citep{2002cra..book.....S}, i.e. long-wavelength perturbations with scales comparable to the particle's Larmor radius $r_L=v\gamma /\Omega_c$, where $\Omega_c = Z eB/m c$ is the cyclotron pulsation of a particle of velocity $v$, mass $m$, and charge $Z$ in a magnetic field of strength $B$. The modeling of particle transport in MHD turbulence is complex. It requires approximations to obtain the solutions that describe the particle trajectories. Transport studies were originally based on the quasi-linear theory (QLT), \citep{1966ApJ...146..480J, 2002cra..book.....S} where the unperturbed (gyromotion) particle trajectory is retained to derive the electromagnetic correlation tensor, itself used to derive the random Lorentz force exerted on the particles. The QLT is applicable over restricted timescale ranges, i.e. intermediate between pitch-angle particle scattering and particle distribution isotropization times. The main drawback of the QLT is that it leads to infinite cosine pitch-angle (the angle between the particle velocity and the background magnetic field) diffusion coefficients at 90 degrees and hence produces a pathological behavior in the calculation of the particle parallel mean free path (mfp). Several analytical attempts have been proposed to deal with this issue (see \cite{2009ASSL..362.....S} and references therein). In particular, the 90 degrees scattering problem has been treated by the mean of the broadening of the resonance \citep{1973Ap&SS..25..471V}. However, the way the resonance is broadened and the non-linear solutions which have to be retained in diffusion coefficient calculations, are model dependent. Needless to say that all these studies are performed in the test-particle approximation (see \cite{2006ApJ...642..902P}). It is known that if the density in the relativistic particles is high enough, the particles can generate their own waves and induce a modification of the turbulence \citep{2004ApJ...604..671F, 2008ApJ...673..942Y, 2011ApJ...731...35Y}. \\
Another difficulty in CR transport studies lies in the description of the random Lorentz forces. In the incompressible MHD limit the turbulence can be described by an anisotropic model \citep{1995ApJ...438..763G, 2001ApJ...554.1175M}. In this model, owing to the Alfv\'en wave packet dynamics, the turbulent perturbations are elongated along the magnetic field and follow a particular scaling relation based on the Kolmogorov phenomenology given by $\bar{k}_{\parallel} \ell^{1/3}= \bar{k}_{\perp}^{2/3}$ (hereafter GS scaling) \footnote{$\bar{k}$ and $k$ are the perturbed wave numbers defined along the local magnetic field $\vec{B}_L$ and the total magnetic field $\vec{B}_T$, respectively. We also define the global or background magnetic $\vec{B}_0$, i.e. the large scale background magnetic field. The total magnetic field includes the contribution of the background magnetic field and the perturbed magnetic field $\delta \vec{B}$. The local magnetic field is not unique. Its direction varies with the scale $1/\bar{k}_{\parallel}$ under specification (see the discussion in \cite{2000ApJ...539..273C}). $\ell$ is the coherence length of the turbulence.}. This relation relies on a critical balance between the perpendicular cascade and Alfv\'en wave packets crossing timescales. In the compressible limit, the situation is more complex: Alfv\'en and slow-magnetosonic modes have been found to follow the same GS scaling, but fast-magnetosonic modes have been found to follow a Kraichnan spectrum and to have an isotropic cascade \citep{2003MNRAS.345..325C}.\\
The impact of the different turbulence models over the CR transport has been discussed in a long list of articles. Apart from the above-mentioned analytical calculations, another approach based on numerical simulations uses synthetic turbulence models and then reconstructs the different transport coefficients by averaging CR trajectories over a large number of particles and magnetic cube realizations \citep{1999ApJ...520..204G, 2002PhRvD..65b3002C, 2004JCAP...10..007C, 2006A&A...453..193M, 2010CoPhC.181...71T, 2011ICRC...10..240S, 2012ApJ...749..103L, 2015JGRA..120.4095H}. In all cases, these calculations impose a model for the turbulent spectrum necessary to derive the particle mfps. With the progress of computational power, some studies have been performed to test CR trajectories in more realistic (or less model-dependent) situations where the turbulence spectrum is calculated by directly solving some fundamental equations. This is particularly the case in the context of CR acceleration at non-relativistic (and also relativistic) shocks where particle-in-cell (PIC) techniques calculate the electromagnetic field solutions from the Maxwell equations. Recent PIC simulations have started self-consistent investigations of particle injection and acceleration in the relativistic regime at supernova remnant shock fronts \citep{2014arXiv1412.1087B}. At higher energies (and larger scales), calculations coupling MHD and PIC techniques have permitted insights into the dynamics of the shock CR precursor \citep{2008MNRAS.386..509R}.\\
In this work, we also use a PIC-MHD approach to investigate the propagation of CRs in magnetized turbulent media. The turbulence is produced by direct numerical simulations using the {\tt RAMSES} MHD code. We follow a few pioneering calculations performed in the context of space plasmas \citep{2012ApJ...750..150W} or interstellar medium studies \citep{2011ApJ...728...60B, 2013ApJ...779..140X}. We first summarize the results obtained in the latter two works as we are more involved in CR propagation in the interstellar medium. We then extend their calculations to include other turbulence regimes and discuss issues connected with the turbulence forcing process. This work presents an investigation of CR transport in MHD turbulence over a large parameter space of Alfv\'enic Mach numbers and particle rigidities. Our simulations, however, are restricted to sub-Alfv\'enic turbulence cases. We note that in this work the Afv\'enic Mach number $M_a$ is defined as the ratio of the rms turbulent fluid velocity $V = \langle \delta V^2 \rangle^{1/2}$ to the Alfv\'en speed taken with the total magnetic field $V_a= B_T/\sqrt{4\pi \rho}$ with $B_T = \sqrt{\delta B^2+ B_0^2}$ and where $\rho$ is fluid mass density. Hence hereafter $M_a < 1$, but also $\delta B/B_0 \le 1$ so that we can always define a parallel and a perpendicular transport with respect to the background magnetic field $\vec{B}_0$.\\

This paper is organized as follows. Section \ref{S:KMHD} presents the general numerical framework adopted in this paper. We describe in \S \ref{S:MHD} the MHD simulations and the forcing procedure as well as the set-up of our simulations. In \S \ref{S:Kinetic} we describe the kinetic module developed to integrate the particle trajectories. Section \S \ref{S:CRMHD} presents the main results obtained in this paper; in \S \ref{S:CRMA} the CR mfp dependence with the Alfv\'enic Mach number is described, while in \S \ref{S:CRRHO} the dependence with particle rigidities is investigated. In \S \ref{S:Conc} we summarize and conclude our work.

\section{Kinetic-MHD simulations}
\label{S:KMHD}
In this work particle trajectories are calculated by means of direct numerical simulation (DNS) of particle transport in MHD turbulence. To this end we upgraded the {\tt RAMSES} MHD code \citep{2002A&A...385..337T, 2006A&A...457..371F} by including a turbulence forcing module (see \S \ref{S:MHD}) and a kinetic module (see \S \ref{S:Kinetic}). 
\subsection{Magneto-hydrodynamic simulations}
\label{S:MHD}
\subsubsection{Forcing module}
\label{S:Force}
The MHD simulations were performed in a 3D cartesian grid with periodical boundary conditions. The HLLD Riemann solver for ideal MHD is used in this work. The simulation box has a size $L=1$ that can be rescaled afterward to define the scale lengths of the problem under consideration. The turbulence is generated by forcing the fluid velocity component with an external force $\vec{f}$ in the Euler equation. The force is expressed in terms of its Fourier transform $\hat{f}$. It is obtained by the realization of independent random Ornstein-Uhlenbeck processes for each excited wave number.  A detailed description of the implementation of the forcing is described in \citet{2009A&A...494..127S} but we reproduce it here for convenience.\\
The components of the force are expressed in terms of the Fourier amplitudes of $N_m$ modes (typically  $N_m=32$). One advantage of the Ornstein-Uhlenbeck process is that each mode component evolves statistically independently from the others. We have
\beq
f_i=\Sigma_{m=1}^{N_m} \hat{f}_{i,m} \cos(2\pi k_{j,m} x_j) \ ,
\eeq
where $\hat{f}_{i,m}$ are initialized to zero at the beginning of each MHD run. Each mode follows a stochastic differential equation:
\beq
\label{Eq:OU}
d\hat{f}_{i,m} = g_{\chi} \times \left[-\bar{\epsilon} \hat{f}_{i,m} {dt \over T} + \bar{\beta} {c_{s} \over T} \sqrt{{2 w(k)^2 \over T}} P^{\chi}_{i,j,m} d\xi_j\right] \ .
\eeq
The different parameters entering in the forcing are $w$ the amplitude of the forcing, $T$ the auto-correlation timescale and $\bar{\epsilon}$ and $\bar{\beta}$ are parameters controlling the relative strength of the deterministic and stochastic forcing components ($\bar{\epsilon}=\bar{\beta}=1$ is adopted in this work) and $d\xi_j = \xi_j \sqrt{dt}$. The random variable $\xi_j$ is sampled over a Gaussian distribution with zero mean and variance one. The tensor $P^{\chi}_{i,j,m}$ defines the geometry of the forcing \citep{2008ApJ...688L..79F, 2011PhRvL.107k4504F},
\beq
\label{Eq:TOU}
P^{\chi}_{i,j,m}=\chi \delta_{i,j} + (1-2\chi) {k_i k_j \over k^2} \ .
\eeq
where $\chi \in [0,1]$ controls the relative importance of solenoidal and compressive modes in the random forcing operator entering in the Ornstein-Uhlenbeck process. Hence, $\chi =0$ corresponds to a pure compressive forcing (hereafter CoF or curl-free) and $\chi =1$ corresponds to a pure solenoidal forcing (hereafter SoF or divergence-free). We also use a mixed forcing (hereafter MoF) with $\chi =0.5$. The normalization factor $g_{\chi} =3/\sqrt{1-2\chi+3\chi^2}$. The fraction of compressible modes generated by a forcing of strength $\chi$ in 3D is derived from the norm of each part of the forcing tensor (right-hand side terms in Eq. \ref{Eq:TOU}). One gets $r_c= F_{comp}/(F_{sol}+F_{comp})= (1-\chi)^2/(1-2\chi+3\chi^2)$ \citep{2010A&A...512A..81F}. Hence $r_c=0, 1/3, 1$ for $\chi=1, 0.5, 0$ respectively. \\
\cite{2004RvMP...76..125M} have proposed different sources of interstellar turbulence: galactic spiral density shocks, large-scale gravitational contraction, supernova explosions, and proto-stellar jets and winds. All are likely to excite compressible modes. Incompressible forcing can, for instance be produced via shearing flows. It is therefore not unrealistic to consider that the turbulence forced at the largest scales has a mixture of compressible and incompressible components. The effect of compressible and incompressible forcing over MHD turbulence has also been addressed in \cite{2013JPlPh..79..597W}. \\

In all of our simulations the isothermal approximation for the gas equation of state is used. The pressure and the density in the simulation set-up are $P=\rho =1$ unless otherwise specified and the sound speed is $c_{s}= \sqrt{\gamma} \simeq 1.005$. The energy is injected into the large-scale modes in the wave number interval $k \in ]1,3] \times 2\pi/N$ (w has a Heaviside function shape in k), where $N$ is defined by the level of refinement $X$ as $N=2^X$. We note that hereafter we make the distinction between $L$, which is the size of simulation cube, and $L_{inj} < L$, which is the scale of turbulence injection identified with the coherence length $\ell$ of the turbulence. 

\subsubsection{Simulation set-up}
We performed different types of MHD simulations; they are summarized in table \ref{T:MHD}.  The nomenclature follows from the value of the forcing parameter $\chi$ and the resolution level $X$. For instance, a simulation at a resolution $N= 512^3 = 2^9$ is at a level X=9. We then specify $\chi$ as $c=0, 0.5, 1$ and $J$ the job number to differentiate jobs with identical $\chi$ values at a given level. We performed simulations at levels 8, 9 and 10. Also displayed in table \ref{T:MHD} are the resulting mean sonic and Alfv\'enic Mach numbers obtained after the turbulent spectrum has reached a quasi-stationary state. The value of the mean Alfv\'enic and sonic Mach numbers are averaged over three snapshots which will be used as turbulent field realizations for the propagation of particles. Each snapshot is separated by several cascading timescales at large scales (i.e. $\sim 10 L_{inj}/V$, where V is the perturbation velocity). The MHD simulations have been performed at the CINES center on the Jade and Occigen super-calculators. The typical duration of a MHD run at $256^3$, $512^3$, and $1024^3$ are 5000, 20000, and 90000 h.cpu, respectively.
\begin{table}
\centering                          
\begin{tabular}{c c c c c c c}        
\hline\hline                 
Name & Level & $w$ & $\chi$ & $\langle \eta \rangle$ & $\langle M_s \rangle$ & $\langle M_a \rangle$ \\    
\hline                        
8J01c0.0 & 8 & 0.04 & 0.0& 0.09 & 0.08  & 0.90    \\
8J02c0.5 & 8 & 0.05 & 0.5& 0.35 & 0.58  & 0.58    \\
8J03c1.0 & 8 & 0.1 & 1.0& 0.83 & 0.1 & 1.00    \\
\hline                        
9J01c0.0 & 9 & 0.04 & 0.0 & 0.25 & 0.35  &0.34  \\
9J02c0.0 & 9 & 0.07 & 0.0 & 0.38 & 0.50  &0.50  \\
9J03c0.0 & 9 & 0.09 & 0.0 & 0.50 & 0.61  &0.66  \\
9J04c0.0 & 9 & 0.18 & 0.0 & 0.71 & 0.92  & 0.88 \\
\hline                        
9J05c0.5 & 9 & 0.02 & 0.5 & 0.18 & 0.37  & 0.37  \\
9J06c0.5 & 9 & 0.045 & 0.5& 0.34 & 0.58  & 0.58 \\
9J07c0.5 & 9 & 0.08 & 0.5 & 0.52 & 0.68  & 0.67 \\
9J08c0.5 & 9 & 0.15 & 0.5 & 0.73 & 0.92  & 0.86 \\
\hline                        
9J09c1.0 & 9 & 0.01 & 1.0 & 0.18 & 0.34  & 0.34 \\
9J10c1.0 & 9 & 0.0225 &1.0& 0.32 & 0.50  & 0.50 \\
9J11c1.0 & 9 & 0.045 & 1.0& 0.49 & 0.69  & 0.66 \\
9J12c1.0 & 9 & 0.1 & 1.0  & 0.72 & 1.0   & 0.88 \\
\hline                                   
10J01c0.5 & 10 & 0.045 & 0.5 & 0.35 & 0.53  & 0.53 \\
10J02c1.0 & 10 & 0.01  & 1.0 & 0.18 & 0.37  & 0.37 \\
\hline                                   
\end{tabular}
\caption{MHD simulations used in this work}             
\label{T:MHD}
\end{table}

Before presenting our calculations in detail we want to clarify an issue, barely explicited, concerning the definitions of the Alfv\'enic Mach number in the literature. In this work we use $M_a= V/V_a$ with $V_a$ defined in the total magnetic field (see \S \ref{S:INT}) and V is the rms turbulent fluid velocity. Instead, several other works use $M_{a0}=V/V_{a0} > M_a$ with $V_{a0}$ defined in the background magnetic field. The turbulence is usually forced in the velocity space so there is additional step to link $M_a$ (or $M_{a0}$) to the level of magnetic fluctuations. Here we consider $\eta=\delta B/\sqrt{(\delta B^2+B_0^2)}$ whereas in many other works $\eta_0=\delta B/B_0$ is used. Table \ref{T:MA} gives the correspondence between $\eta$ and $\eta_0$. The advantage of using $\eta$ is that this quantity is bounded between 0 and 1. In our work both $\eta$ and $M_a$ are averaged over three MHD snapshots and are written $\langle \eta \rangle \le \langle M_a \rangle$. In table \ref{T:MHD} the value of $\langle \eta \rangle$ for the different jobs is given in the fifth column. Comparing these values with those dispalyed in table \ref{T:MA} we see that our simulations are in the sub- to trans-Alfv\'enic regime with respect to the Alfv\'en velocity taken in the background magnetic field.

\begin{table}
\centering                          
\begin{tabular}{|c|c|}        
\hline\hline                 
$\eta$ & $\eta_0$  \\    
\hline                        
0. & 0.     \\
0.10 & 0.10     \\
0.30 & 0.31     \\
0.50 & 0.57     \\
0.60& 0.75     \\
$1/\sqrt{2} \sim$ 0.707 & 1.00  \\
 0.80 & 1.33 \\                                                                                        
0.90 & 2.10\\
1.00 & $\infty$ \\
\hline                        
\end{tabular}
\caption{Correspondence between the parameters $\eta$ and $\eta_0$.}             
\label{T:MA}                                                                                                                  
\end{table}

\subsection{Kinetic module}
\label{S:Kinetic}
\subsubsection{General description}
We solved the Lorentz equation for each particle of charge $q$ and mass $m$. The particle has a momentum $\vec{p} = \gamma m \vec{v}$ and a normalized velocity $\vec{\beta}=\vec{v}/c$  and propagates in an electromagnetic field $\delta \vec{E}$ (no mean electric field), $\vec{B}_T = \delta{\vec{B}}+\vec{B}_0$:
\bea
\label{Eq:Lorentz}
{d\vec{p} \over dt} &=& q \delta \vec{E} + q \vec{\beta} \wedge \vec{B}_T  \ , \nonumber \\
{d\vec{r} \over dt} &=& \vec{v} \ .
\eea
The electromagnetic fluctuating components $ \delta \vec{E}$ and $\delta{\vec{B}}$ are provided by the MHD code. In this work, only protons are considered, hence $m=m_p$ and $q=+e$. Each particle is injected with a Lorentz factor $\gamma_0$. This defines the particle's Larmor radius $r_{L0}=\gamma_0 mc^2/eB_0$. We also define the particle's synchrotron pulsation $\Omega_0=r_{L0}/c$. Eq. \ref{Eq:Lorentz} is normalized with respect to this initial value as (see \cite{2011ApJ...728...60B}, hereafter BYL11):
\bea
\label{Eq:LorentzB1}
{d \hat{u} \over d \hat{t}} &=& \hat{\gamma} {\delta \vec{E} \over B_0} + \hat{u} \wedge ({\delta \vec{B} \over B_0} + \vec{e}_z) \ , \\
\label{Eq:LorentzB2}
{d \hat{x} \over d \hat{t}} &=& {r_{L0} \over L} \times \hat{u} \ .
\eea
We assume that the large-scale magnetic field $\vec{B}_0$ is aligned along $\vec{e}_z$. In the previous equations we have the following notations: $\hat{u} = \hat{\gamma} \vec{\beta}$, $\hat{\gamma}= \gamma/\gamma_0$, $\hat{t} = t \times (\hat{\gamma} r_{L0}/c)^{-1}$. In practice, at least in the interstellar medium, the effect of the electric field over high-energy (multi TeV) CRs can be neglected and $\hat{\gamma}$ remains equal to 1.\\
When particles are submitted to radiative losses (or re-acceleration) it is necessary to add another equation for $\hat{\gamma}$. For instance, relativistic protons cooling under the effect of pion production with a cross section $\sigma_{pp}$ verify
\[
{d \ln{\hat{\gamma}} \over d\hat{t}} =- \hat{\gamma} \times {r_{L0} \over c t_{pp}} \ ,
\]
where $t_{pp} \simeq (\sigma_{pp} n_H c)^{-1}$ for a medium of hydrogen density $n_H$. We note that hereafter, as the particle energy is fixed, we denote $r_{L0}$ and $\Omega_0$ to $r_{L}$ and $\Omega$.

\subsubsection{Integration schemes}
We tested several integration schemes for Eqs. \ref{Eq:LorentzB1} and \ref{Eq:LorentzB2}: a leap-frog scheme (by default implemented in {\tt RAMSES} to treat the propagation of test particles in a gravitational field), a Runge-Kutta method of 5th order, and a Bulirsch-Stoer method (\cite{1992nrfa.book.....P}). We tested the different integration methods and found differences at the level of a few percent over the particle mfps. This small difference can be explained by the construction of the code: {\tt RAMSES} adapts the time step in the PIC module in a way that a particle cannot cross more than one grid cell within one time step (see \cite{2002A&A...385..337T}, \S 2.4). We used the leap-frog second-order integration method to save computational resources. The essential effect now is the magnetic field calculation at the particle position (see next section). The typical duration of one run to complete the propagation of $10^6$ particles depends on the particle's rigidity and the Alfv\'enic Mach number. It varies from $10^6$ to a few $10^7$ time steps, in other words from ten hours to a few tens of hours of computation on the Jade and Occigen supercomputers at the CINES facility.

\subsubsection{Electromagnetic field interpolation}
The magnetic field is interpolated at the position of the particle from its values derived by the MHD code at the grid points. For this, we have adopted a volume-averaged interpolation of the field strength. The volume can include the next eight neighbor points, this is the first-order cloud-in-cell (CIC) interpolation implemented by default in {\tt RAMSES} or the next 64 neighbor points using a third-order piecewise cubic spline (PCS) interpolation (\cite{2013PhPl...20f2904H}). The results in this work were obtained using the PCS interpolation method.

\section{Direct numerical simulations of cosmic ray transport}
\label{S:CRMHD}
Direct numerical simulations of CR transport in magnetic turbulence is a young but already extended field of research. In \S \ref{S:Sum} we summarize the results of two related works proposed by BYL11 and \cite{2013ApJ...779..140X} (hereafter XY13), both developed in the context of CR propagation in the interstellar medium. In \S \ref{S:CRMA} we present our results describing CR mfp dependences with respect to Alfv\'enic Mach numbers while in \S \ref{S:CRRHO} we present our results describing CR mfp dependences with respect to CR rigidities. In \S \ref{S:FOR} we discuss how forcing geometries affect the CR transport. Finally \S \ref{S:RAT} presents the variations of the ratio of perpendicular to parallel mfps with the particle rigidity at different turbulence levels.

\subsection{Summary of previous works}
\label{S:Sum}
\cite{2011ApJ...728...60B} have investigated the propagation of CRs in incompressible forced MHD turbulence. In the incompressible limit, the effects of the fast-magnetosonic modes are absent, the transport is controlled by the shear-Alfv\'en and the pseudo-Alfv\'en modes (the incompressible limit of slow-magnetosonic modes). Both types of modes have been found to follow a GS scaling \citep{2001ApJ...554.1175M}. BYL11 solve Eqs. \ref{Eq:LorentzB1} and \ref{Eq:LorentzB2} neglecting the effect of electric field fluctuations and use different statistically independent (static) magnetic configurations. No back-reaction of CRs over the turbulence is considered in this work. \\
\cite{2011ApJ...728...60B} have conducted 3D simulations at a resolution of $768^3$ in both balanced and imbalanced turbulence cases. Imbalanced turbulence is expected, for instance close to CR sources where a gradient of CR can drive perturbations in a preferred direction with respect to the mean magnetic field or in the solar wind. The turbulence is injected at large scales using a stochastic solenoidal forcing (see the definition in \S \ref{S:Force}). The authors have carried out simulations in a regime of weak turbulence (or strong background field) with an Alfv\'enic Mach number $M_a = 0.1$ and in a regime of strong turbulence (or weak background field) with $M_a =1$. We note that BYL11 calculate the Alfv\'en velocity with respect to the background magnetic field $B_0$, hence $M_a$ has to be interpreted as $M_{a0}=\delta B/B_0$ in our work. The injection scale of the turbulence in their simulation is (in box length units) $L_{inj} \simeq 0.2 L$. We note that in the sub-Alfv\'enic case, BYL11 use an elongated box with a main axis parallel to $\vec{B}_0$. Then, in that case $L_{inj}$ corresponds to the perpendicular turbulent outer-scale and the parallel outer-scale of the turbulence in the parallel direction is $10 L_{inj}$. \\
\cite{2011ApJ...728...60B} have computed the variation of spatial diffusion coefficients with respect to the direction of the background magnetic field $\vec{B}_0$. Balanced and imbalanced turbulence cases do not show strong differences and result in diffusion coefficients varying by a factor of less than a few. In the case of balanced trans-Alfv\'enic turbulence, BYL11 find a parallel diffusion coefficient $D_{\parallel}$ independent of the particle's rigidity at low rigidity $\rho=r_L/L \le 0.02$ \footnote{We adopt the following notations: $\tilde{\rho}$ is the reduced rigidity with respect to the turbulence outer scale and $\rho$ is the reduced rigidity in cube units; in BYL11 and in our work we have $\tilde{\rho} \sim 5 \rho$.} then scaling as $\rho^{1/3}$ in the range $0.02 <\rho < 0.1$ and scaling as $\rho^2$ beyond (see their figure 6). Considering perpendicular diffusion, the authors claim to have a coefficient independent of the particle's rigidity in both sub-Alfv\'enic and trans-Alfv\'enic cases. In fact, considering their figures 7 and 8, their solutions show stronger dependence on rigidity. Especially, $D_{\perp} \propto \rho^{1/3}$ fits the curve corresponding to the balanced turbulence case in the trans-Alfv\'enic case better. We find the same trend in the sub-Alfv\'enic case for $\rho \le 0.05$. No explicit effects about the turbulence level has been reported, but a gain of a factor of $\sim 3-4$ of difference for $D_{\perp}$ can be deduced between sub- and trans-Alfv\'enic cases.\\

\cite{2013ApJ...779..140X}  have investigated the propagation of CRs in compressible MHD turbulence and so the simulations retain the effects of shear-Alfv\'en modes and both magnetosonic modes. Here again the magnetostatic and test-particle limits have been retained. \\
\cite{2013ApJ...779..140X}  conducted 3D simulations at a resolution of $512^3$. The turbulence is injected at large scales using a stochastic forcing of the velocity field. The forcing is purely solenoidal. The authors carry simulations in a regime of weak turbulence with Alfv\'enic Mach numbers ranging from $M_a = 0.19$ to $M_a=0.73$. XY13 have calculated the Alfv\'en velocity with respect to the total magnetic field $B_T$ hence their definition of $M_a$ is similar to our. The injection scale of the turbulence in their simulation is (in box length units) $L_{inj}=0.4 L$. \\
\cite{2013ApJ...779..140X} computed spatial diffusion coefficients with respect to the direction of the background magnetic field $\vec{B}_0$. The authors conducted a detailed study of the effect of the turbulence level over the particle transport and have calculated CR parallel and perpendicular mfps dependences with respect to $M_a$. The results, however have been presented only at one particle rigidity $\rho = 0.01$ (so in box length units). Parallel mfps have been found to be $> L_{inj}$ scaling roughly as $M_a^{-2}$. Perpendicular mfps have been found to be consistent with a $M_a^4$ scaling expected theoretically by \cite{2008ApJ...673..942Y} in the case the parallel mfps verify $\lambda_{\parallel} > L_{inj}$ (see details in \S \ref{S:DISMA}).\\ 

We present our results below. We provide an extensive investigation of CR parallel and perpendicular mfps dependences with respect to $M_a$ and $\rho$ that extend the two above works significantly. We find power-law variations for $\lambda_{\parallel}$ and $\lambda_{\perp}$. The power-law index is denoted $\alpha$ hereafter. 

\subsection{CR mean free paths: Calculation procedure}
Cosmic ray mfps are reconstructed using an averaging procedure over a large number of individual particle trajectories. We record the position $x,y,z$ in the 3D simulation cube for each particle: $z(t)$ determines the position of the particle with respect to the background magnetic field $\vec{B}_0$, $x(t),y(t)$ determines the position of the particle in the directions perpendicular to the background magnetic field $\vec{B}_0$. \\
To proceed to mfp calculations, we have selected a set of $N_r=3$ magnetic field realizations separated by at least two large-scale cascade times in order to ensure a statistical independence between two successive realizations. For each realization an ensemble of $N_p$ particles are propagated until a convergence is obtained (see next). The parallel spatial diffusion coefficient is calculated using the following formula:

\beq
\kappa_{\parallel}(t) = {\langle(z(t)-z(0))^2\rangle \over 2 t} = {1 \over N_r} {1 \over N_p} \Sigma_{i=1}^{N_r} \Sigma_{j=1}^{N_p} {(z_{i,j}(t)-z_{i,j}(0))^2 \over 2 t} \ .
\eeq

The perpendicular spatial diffusion coefficient is calculated using the following formula:

\beq
\kappa_{\perp}(t) = {1 \over N_r} {1 \over N_p} \Sigma_{i=1}^{N_r} \Sigma_{j=1}^{N_p} {(x_{i,j}(t)-x_{i,j}(0))^2+ (y_{i,j}(t)-y_{i,j}(0))^2 \over 4 t} \ .
\eeq

\begin{figure}
\centering
 \includegraphics[width=6cm, angle=-90]{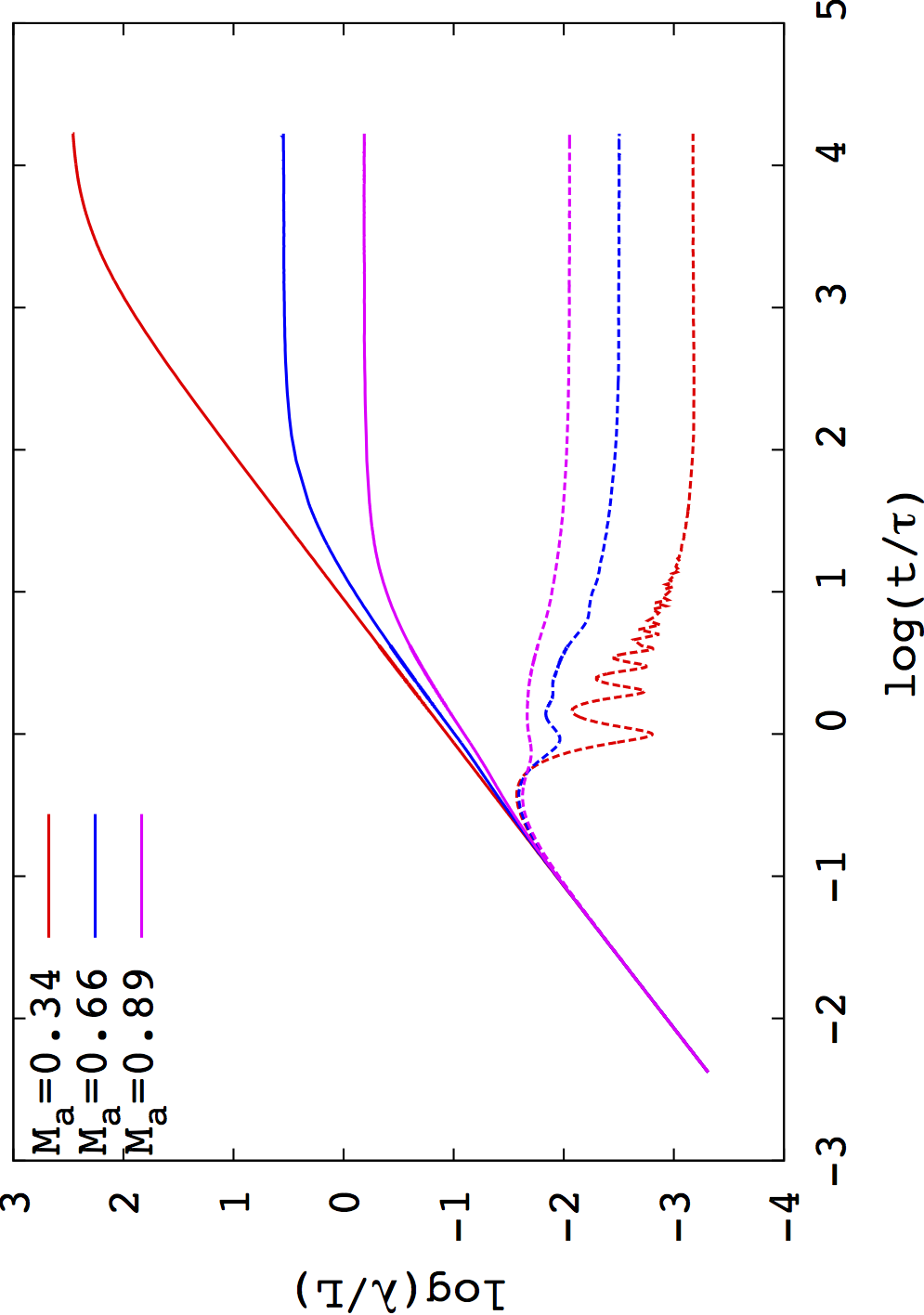}
 \caption{Time-dependent evolution of $\lambda_{\parallel}(t)$ (continuous lines) and  $\lambda_{\perp}(t)$ (dashed lines) for a normalized particle Larmor radius $\rho=r_L/L=0.037$ at three different Alfv\'enic Mach numbers with $\chi=1$ (jobs 9J09c1.0, 9J11c1.0 and 9J12c1.0). The mfp corresponds to the plateau reached at large timescales. Time is in units of particle gyroperiod $\tau=\Omega^{-1}$.} 
\label{F:Fig1}
\end{figure}

These two coefficients are calculated until they both converge to a plateau (see an example in figure \ref{F:Fig1} for a particle at $\rho=r_L/L=0.037$). We note that at low $M_a$ the ballistic regime is present up to $t \sim 10^3 \Omega^{-1}$. Once the diffusion coefficient $\kappa_{\parallel,\perp} = \lim \limits_{t \rightarrow +\infty} \kappa_{\parallel,\perp}(t)$ is found the corresponding mfp is obtained using the relation $\lambda_{\parallel,\perp}=3 \kappa_{\parallel,\perp}/v$ (the particle velocity is identified to the light velocity ($v=c$) in the cases considered here). For the results presented below a typical number of $N_p = 10^6$ particles is used. 

\subsection{CR mean free paths: Alfv\'enic Mach number dependences}
\label{S:CRMA}
\begin{figure}
\centering
 \includegraphics[width=7.cm, angle=+90]{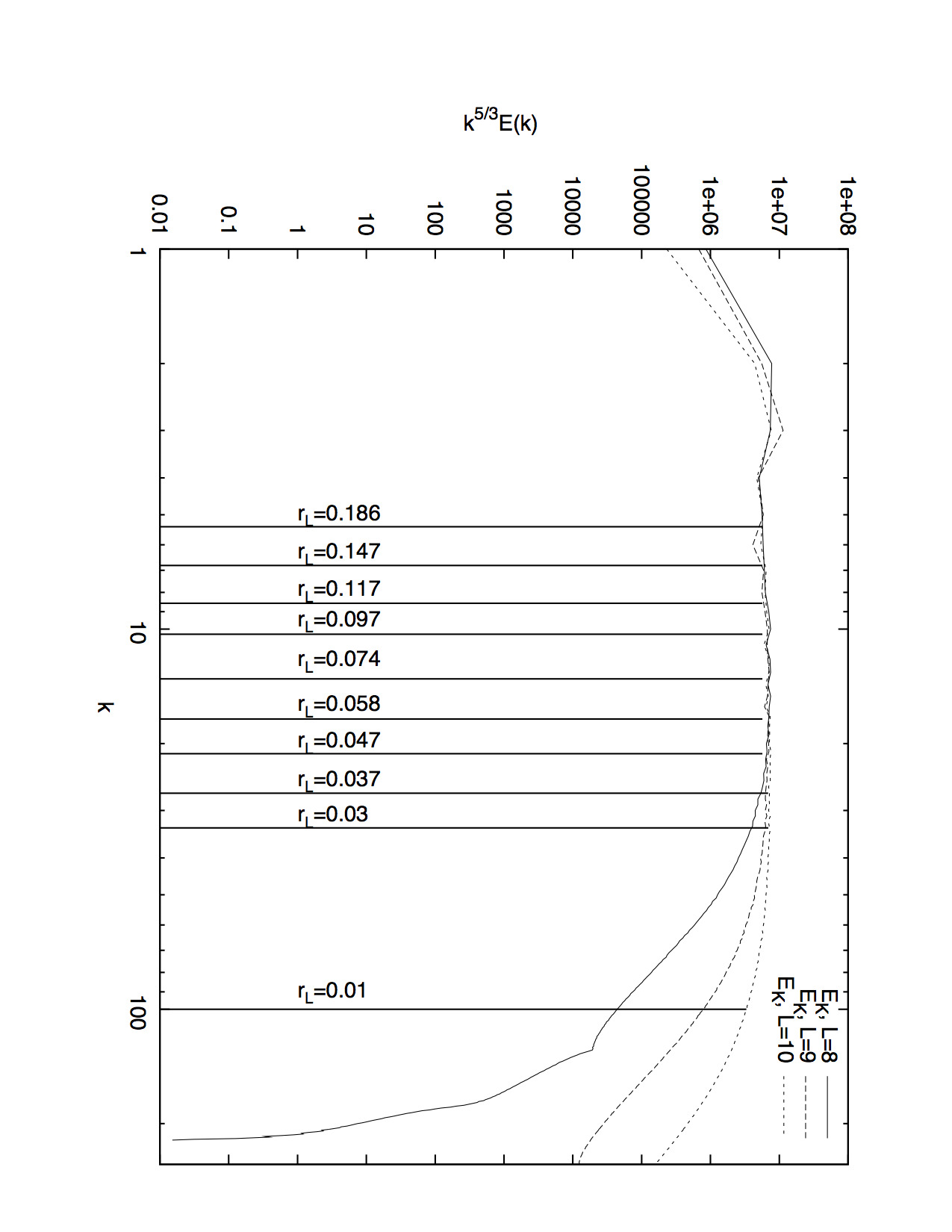}
 \caption{Stationary kinetic energy turbulent spectra corresponding to jobs 8J01c0.5, 9J06c0.5, and 10J01c0.5 in table \ref{T:MHD}. Also displayed: the normalized wavenumbers $k L$ corresponding to different normalized Larmor radii $\rho =r_L/L$.} 
\label{F:Fig2}
\end{figure}

We first present the dependence of $\lambda_{\parallel}$ and $\lambda_{\perp}$ with respect to the turbulence level $M_a$. In the following figures, the selected $\rho$ are $0.01, 0.037, 0.047, 0.058, 0.074, 0.097, 0.117, 0.147$, and $0.186$ respectively. We note that all rigidities except for $\rho=0.01$ can be associated with a scale in the inertial range of the turbulence as can be seen in figure \ref{F:Fig2}. \\
Another important aspect to keep in mind while comparing these results with the transport in the interstellar medium (ISM) or interplanetary medium (IPM) is that our solutions are restricted to high-energy particles. To fix the ideas we consider typical magnetic fields of $5 \mu$G in the ISM and $50 \mu$G in the solar wind. Hence, considering turbulence injection scales expressed in parsec and astronomical units, we have $\tilde{\rho}_{ISM} \sim 0.21 E_{PeV} B_{5\mu G}^{-1} L_{inj,pc}^{-1}$ and $\tilde{\rho}_{IPM} \sim 4.4\times 10^{-3} E_{GeV} B_{50\mu G}^{-1} L_{inj,AU}^{-1}$ in the ISM and IPM, respectively. Here $E_{PeV}$ and $E_{GeV}$ are the particle energies expressed in PeV and GeV units. \\

\subsubsection{Parallel mean free paths}
\begin{figure}
\centering
 \includegraphics[width=6cm, angle=-90]{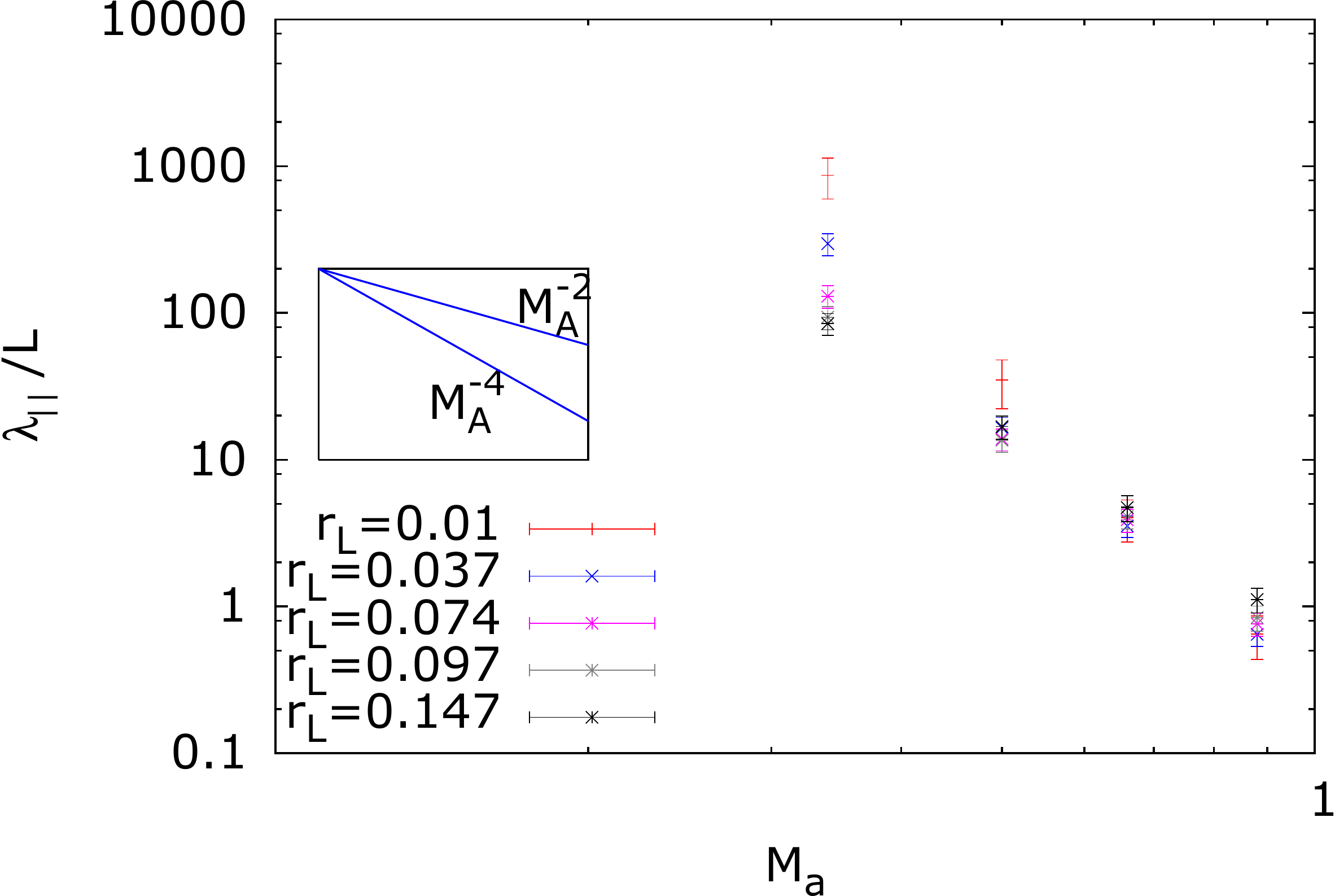}
 \includegraphics[width=6cm, angle=-90]{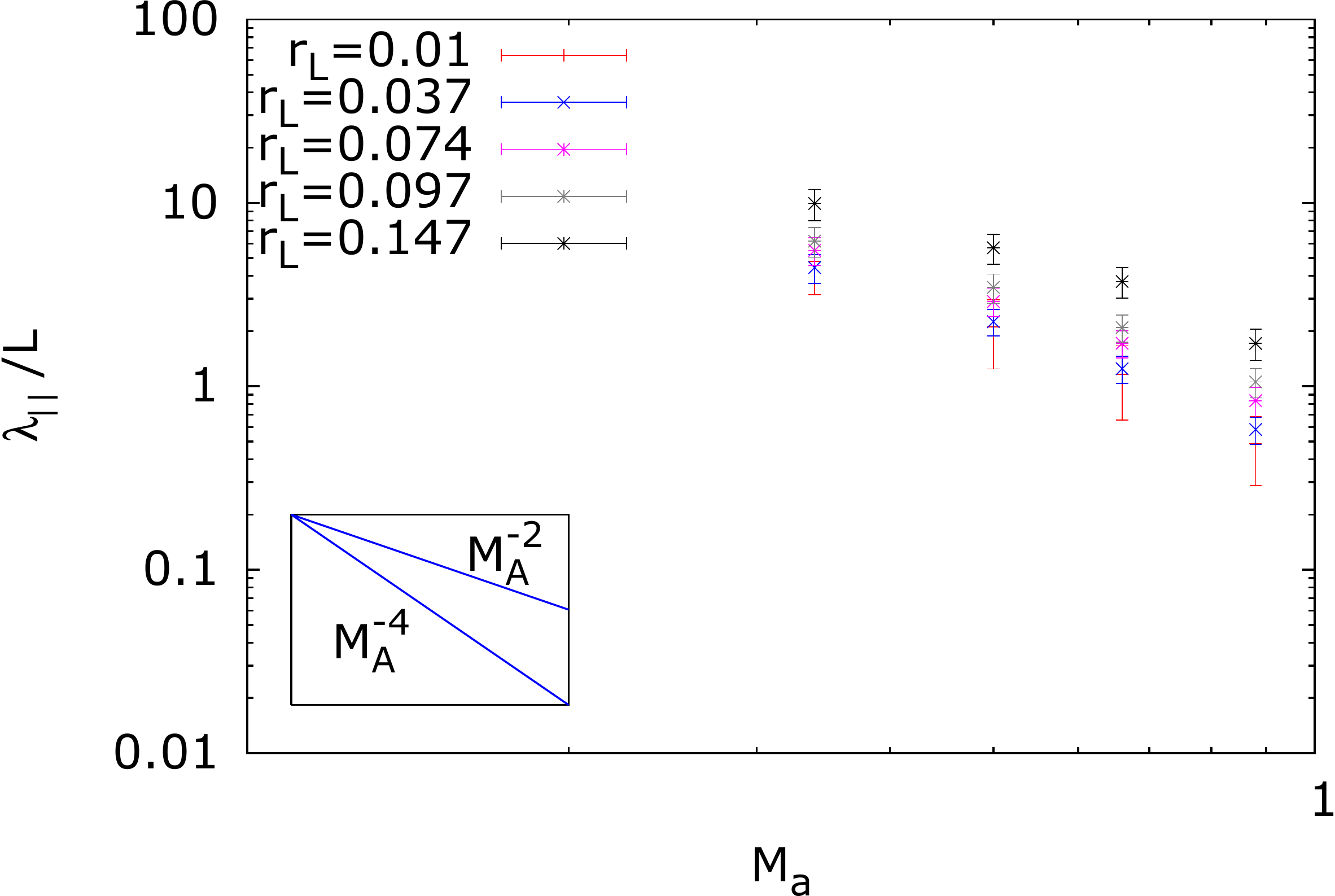}
 \caption{[Upper figure] Parallel mfp versus Alfv\'enic Mach numbers $M_a$ at different particle rigidities in the case of a solenoidal forcing ($\chi =1$). [Lower figure] Parallel mfp versus Alfv\'enic Mach numbers $M_a$ at different particle rigidities in the case of a compressible forcing ($\chi =0$). $M_a^{-2}$ and $M_a^{-4}$ dependences are shown in the insert.} 
\label{F:Fig3}
\end{figure}

In figure \ref{F:Fig3}, we show the dependence of the CR parallel mfps with respect to the Alfv\'enic Mach number (the upper figure displays the results obtained for $\chi=1$ (SoF), the lower figure displays the results obtained for $\chi=0$ (CoF)). It can be seen that parallel mfps decrease with $M_a$ at all particle rigidities. The SoF solutions show faster variations with $M_a$ than the CoF solutions. We find typical ratios $\lambda_{\parallel}(\chi=1)/\lambda_{\parallel}(\chi=0)$ varying between $\sim 100$ at low $M_a$ to $\sim 1$ as $M_a \rightarrow 1$. We find that $\lambda_{\perp}(\chi=1)/\lambda_{\perp}(\chi=0)$ varies from 4-5 for $M_a < 0.6$ to $\sim 1$ as $M_a \rightarrow 1$. In figure \ref{F:Fig3} (the SoF case) it can also be seen that the parallel mfp decreases with an increasing rigidity, which is an unsual behavior. We come back to this issue in \S \ref{S:DISRL}. \\

We extracted a subset of solutions at two different Larmor radius $\rho = 0.01$ and $\rho=0.097$. The former allows us to make some comparisons with XY13, while the latter corresponds to a particle Larmor radius in resonance with a wave number $k_{0.097}$ deep in the inertial range of the turbulence. It corresponds to a scale reasonably far from the injection zone (typically $k_{0.097} L \sim 10, k_{0.097} L_{inj} \sim 2)$, as in our case $L_{inj} \sim 0.2 L$. Particles with rigidities in the inertial range are interesting for investigating the effect of resonant pitch-angle scattering over particle parallel mfps. The difficulty is that resonant interactions occurring at $k r_L \sim 1$ are important in the calculation of $\lambda_{\parallel}$, and non-resonant interaction produced by the transit-time damping (TTD) mechanism due to fluctuations at scales larger than $r_L$ also contribute \citep{1975ApJ...198..485L}. The error bars correspond to standard deviations produced by the solutions from the different snapshots. \\

In the case of incompressible (solenoidal) forcing $\chi =1$ we find the following solutions:
\bea
\label{Eq:LPAX1}
\lambda_{\parallel} &\propto& M_a^{-7.69 \pm 0.57} \ , \rm{\rho=0.01} \nonumber \\
\lambda_{\parallel} &\propto& M_a^{-4.91 \pm 0.35} \ , \rm{\rho=0.097}
\eea
In the case of compressible forcing $\chi =0$ we find the following solutions:
\bea
\lambda_{\parallel} &\propto& M_a^{-2.44 \pm 0.82} \ , \rm{\rho=0.01} \nonumber \\
\lambda_{\parallel} &\propto& M_a^{-1.82 \pm 0.35} \ , \rm{\rho=0.097}
\eea
The error on the index is obtained considering the extreme slopes that fit all data points. The mean index values and their errors depend on the errors produced by each data point. We have compared our results using one, three and six snapshots for the job 9J06c0.5. We found that mfps in each of the last two cases are contained within the error of the simulation at one snapshot. The errors obtained using three and six snapshots are similar to (even if slightly reduced in the case with six snapshots), but are reduced with respect to the case with one snapshot. In order to optimize the averaging procedure we keep the number of snapshots $N_r=3$ in all our calculations.

\subsubsection{Perpendicular mean free paths}
\begin{figure}
\centering
 \includegraphics[width=6cm, angle=-90]{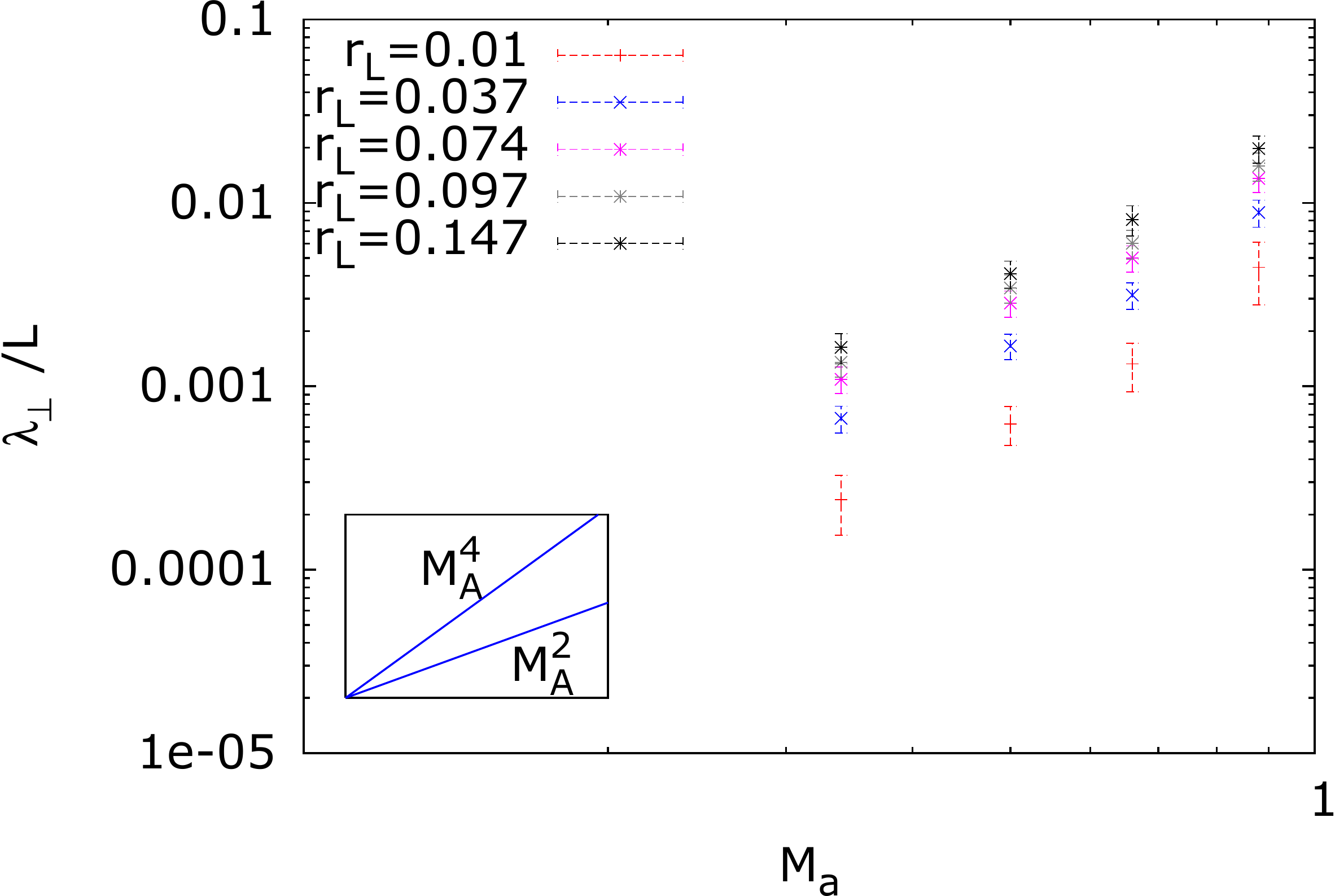}
 \includegraphics[width=6cm, angle=-90]{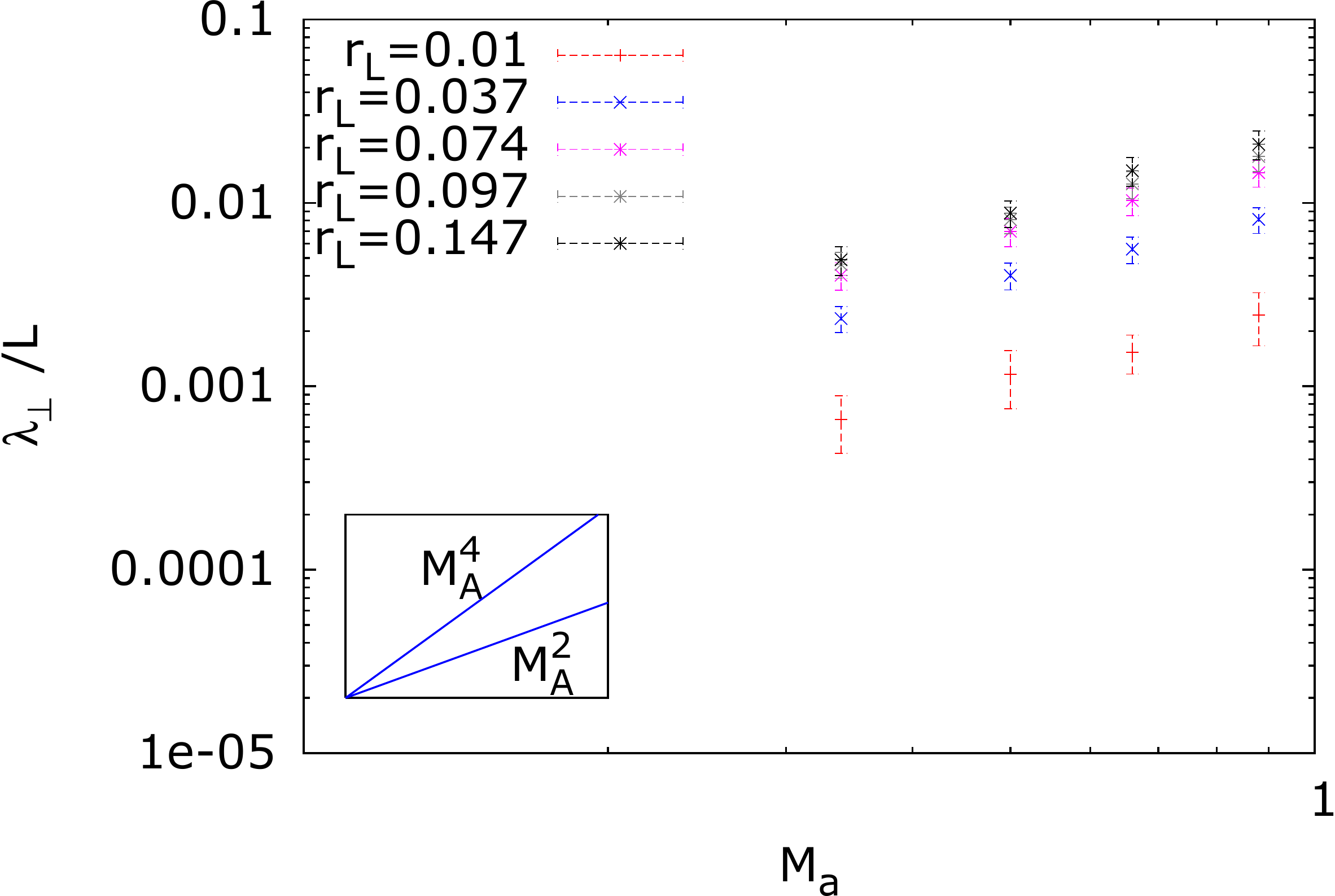}
 \caption{[Upper figure] Perpendicular mfp versus Alfv\'enic Mach numbers $M_a$ at different particle rigidities in the case of a solenoidal forcing ($\chi =1$). [Lower figure] Perpendicular mfp versus Alfv\'enic Mach numbers $M_a$ at different particle rigidities in the case of a compressible forcing ($\chi =0$).} 
\label{F:Fig4}
\end{figure}

The perpendicular mfps are controlled by two efects: a first contribution is associated with the different processes entering in the parallel transport (resonant and non-resonant pitch-angle scattering) and a second contribution is associated with the wandering of magnetic field lines \citep{1971RvGSP...9...27J}. Figure \ref{F:Fig4} shows the CR perpendicular mfps as a function of the Alfv\'enic Mach number.\\

In the case of solenoidal forcing $\chi =1$ we find the following solutions:
\bea
\label{Eq:LPEX1}
\lambda_{\perp} &\propto& M_a^{2.94 \pm 0.69} \ , \rm{\rho=0.01} \nonumber \\
\lambda_{\perp} &\propto& M_a^{2.52 \pm 0.26} \ , \rm{\rho=0.097}
\eea
In the case of compressible forcing $\chi =0$ we find the following solutions:
\bea
\lambda_{\perp} &\propto& M_a^{1.39 \pm 0.73} \ , \rm{\rho=0.01} \nonumber \\
\lambda_{\perp} &\propto& M_a^{1.42 \pm 0.37} \ , \rm{\rho=0.097}
\eea

\subsubsection{Discussion}
\label{S:DISMA}

{\it Comparison with previous works:} We first compare our results in the SoF case ($\chi=1$) with the publication of XY13 where the mfp was calculated at $\rho=0.01$ only. Comparing our figure \ref{F:Fig3} (upper figure, red points) with figure 5 in XY13 we find some discrepancies especially in the regime $M_a < 0.7$ where we find $\lambda_{\parallel}/L > 100$, whereas XY13 has typical mfps $\lambda_{\parallel}/L < 10$. At higher Alfv\'enic Mach numbers the two results are compatible. This effect is due to the much larger index $\alpha \sim -7.7$ obtained in our simulations (see Eq. \ref{Eq:LPAX1}). Our CoF results at $\rho=0.01$ are in good agreements with XY13 solutions. \\
If we now compare the perpendicular mfp results (the red points in our upper figure \ref{F:Fig4} and figure 6 in XY13) we find reasonably good agreement: our index is $\alpha=2.94 \pm 0.69$ and is compatible with the XY13 index $\alpha=4.21 \pm 0.75$. The normalization factor of the mfps is in agreement since we normalize $\lambda_{\perp}$ to the box length $L$, whereas XY13 normalized $\lambda_{\perp}$ to the injection scale of the turbulence $L_{inj}$. If the wandering of magnetic field lines dominates the CR perpendicular transport it is not unrealistic to obtain similar results for $\lambda_{\perp}$ in both studies.\\
At this stage it is difficult to explain the reason for the discrepancy found between the two works for $\lambda_{\parallel}$ at low and moderate $M_a$. What can be said is that even if the forcing is incompressible in both cases, the forcing method is not similar in the two MHD codes. Also, the structure of the two MHD codes and the methods used in the codes to derive fluid solutions are probably different. Another critical aspect is that XY13 derived their solutions at a normalized rigidity $\rho=0.01$. Particles at this rigidity have a Larmor radius that is in resonance with perturbations that are in the dissipation range of the turbulence for simulations at a resolution of $512^3$ (as can be clearly seen in figure \ref{F:Fig2}). Wave-particle resonance is a critical process that controls particle scattering along the magnetic field lines. It can be seen especially in the SoF case (and to a lesser extent in the CoF case) as all rigidities except for $\rho=0.01$ (and also $\rho=0.037$ at low $M_a$) have a uniform dependence on  $M_a$. This is the main reason why with regard to the problem of propagation of CR in MHD turbulence - we suggest that the results (here and in XY13) derived at $\rho=0.01$ have to be interpreted with some care at least concerning $\lambda_{\parallel}$. This is also why hereafter, unless otherwise specified, we concentrate our analysis on Larmor radii that correspond to scales  in the inertial range of the turbulent spectrum (that is to say all rigidities but $\rho=0.01$ in figure \ref{F:Fig2}).\\

\noindent{\it Theoretical frameworks:} QLT predicts $\lambda_{\parallel} \propto M_a^{-2}$ because in this approximation the scattering frequency is $\propto M_a^2$ (see e.g. \cite{2011ICRC...10..240S}) \footnote{\cite{1968PhRvL..21...44J} instead predicted $\lambda_{\parallel} \propto \eta_{0}^{-2}$ and $\lambda_{\perp} \propto \eta_{0}^2$.}. \cite{2002PhRvD..65b3002C} in the case of isotropic turbulence have obtained $\lambda_{\parallel} \propto \eta^{-2} \propto M_a^{-2}$ in the limit $\delta B \ll B_0$. They have found this scaling to be valid in the regime $\eta \rightarrow 1$, i.e. beyond the QLT validity domain (see their appendix A) \footnote{\cite{2002PhRvD..65b3002C} defined $\eta$ as a quantity which corresponds to $\eta^2$ in our text.} . A deviation from the QLT result is also expected in the case of a slab-type turbulence with a steep inertial spectrum index \footnote{This is the 1D wave number index, denoted $s$ in \S \ref{S:DISRL}. Deviations from QLT are obtained for $s>2$.} a result obtained using a second order quasi-linear theory \citep{2009A&A...507..589S}.\\ 
For perpendicular mfps, QLT is obtained in the limit of vanishing scattering (or infinite mfp) along the mean magnetic field (see \cite{1968PhRvL..21...44J, 2011ICRC...10..240S}). In that regime, field line wandering controls the perpendicular CR transport and leads to $\lambda_{\perp} \propto M_a^2$. However, recent theoretical predictions have been proposed by \citet{2008ApJ...673..942Y} (see the discussion in their \S 5) \footnote{\citet{2008ApJ...673..942Y} used $M_{a0}$ instead of $M_{a}$.} in the framework of the compressible MHD turbulence model. Sub-Alfv\'enic turbulence is characterized by two regimes \citep{2006ApJ...645L..25L}: between $L_{inj}$ and $\ell_{tr}=L_{inj} M_a^2$ the turbulence is weak, the magnetic power spectrum develops in the perpendicular direction with respect to the mean magnetic field and scales as $k_{\perp}^{-2}$; beyond $\ell_{tr}$ up to scales such that $k L_{inj} \sim 40-50$, which fall in the dissipative range, GS scaling applies and the turbulence is strong. The CR perpendicular mfp is mostly controlled by the wandering of magnetic field lines but modulated by the effect of scattering: if $\lambda_{\parallel}/L_{inj} > 1$, then perturbations uncorrelated at scales larger than $\ell_{tr}$ produce a perpendicular mfp $\lambda_{\perp}/L_{inj}= 1/3 M_a^4$. If conversely $\lambda_{\parallel}/L_{inj} < 1$, then diffusion along the field lines lead to $\lambda_{\perp}= \lambda_{\parallel} M_a^4$. Finally we note that \cite{2002PhRvD..65b3002C} in the case of isotropic Kolmogorov turbulence numerically obtained $\lambda_{\perp} \propto \eta^{4.6} \lambda_{\parallel}$, i.e. from the above discussion $\lambda_{\perp} \propto \eta^{2.6}$.\\
\cite{2014ApJ...784...38L} explored the effect of chaotic magnetic field behavior over CR transport. The authors confirmed that the CR perpendicular transport is diffusive in the regime $\lambda_{\parallel}/L_{inj} > 1$ for curvilinear distances $s$ along the magnetic field lines larger than a few tens of $L_{inj}$. However, they described the existence of a super-diffusive perpendicular transport regime produced by the strong turbulence regime at intermediary scales $s/L_{inj} \sim 1-10$ (see Eq. 10 in \cite{2014ApJ...784...38L}).\\

\noindent{\it Parallel mean free path:} From the upper part of figure \ref{F:Fig3} it can be seen that our SoF results are incompatible with the predictions of QLT as $\alpha <-2$. Even if we remove the two highest $M_a$ points where QLT is questionable, we still find $\alpha < -2$. We will consider this aspect again in \S \ref{S:DISRL} while discussing the magnetic power spectra in the SoF limit. In the CoF case, we obtain parallel mfps with indices close to -2 and this even including the highest $M_a$ points. This means that our results are consistent with QLT predictions in the low $M_a$ regime and extend it in the high $M_a$ regime. Our CoF results are consistent with the theoretical predictions of \cite{2002PhRvD..65b3002C}.\\
Comparing the upper and the lower parts of figure \ref{F:Fig3}, it is clear that SoF ($\chi=1$ ) simulations produce a turbulent spectrum that leads to a less efficient CR scattering (see also \S \ref{S:DISRL}). The SoF favors the production of incompressible Alfv\'en modes with respect to compressible (in particular fast-magnetosonic) modes even if both types of modes are driven at large scales. At smaller scales, the production of fast modes from Alfv\'en modes has been found to be weak at low levels of turbulence \citep{2003MNRAS.345..325C}. One can expect that the SoF solutions at low $M_a$ provide a good hint of the effect of Alfv\'en waves over the CR transport. Alfv\'en waves at the lowest perturbation order do not contribute to TTD but only to Landau-synchrotron resonance (also known as gyro-resonance), so parallel mfps in the SoF case likely result from this process. Recent transport models (see e.g. \cite{2000PhRvL..85.4656C}) concluded that Alfv\'enic turbulence does not produce an efficient CR scattering because of the strong anisotropy the perturbations develop towards small scales. At these scales, the resonant condition implies that within a Larmor radius a particle interacts with several uncorrelated perturbations resulting in an inefficient pitch-angle scattering. As reported above, we find typical ratios $\lambda_{\parallel}(\chi=1)/\lambda_{\parallel}(\chi=0) \sim 10-100$ at low $M_a$ depending on the particle rigidity. This partly supports the conclusion of the above analysis, but in the mean time the ratio is not as small as predicted by this model. \cite{Yan02} have improved Chandran's model by considering an Alfv\'enic magnetic tensor derived from numerical calculations of \cite{2003MNRAS.345..325C} and found a parallel mfp larger by $\sim$ 4 orders of magnitude above $\sim 1 \rm{GeV}$, which is closer to our results. It is also possible that a fraction of the energy of the driven velocity field is transferred into fast modes and/or that the gyro-resonance effect in Afv\'enic turbulence is stronger than expected. For $\rho <0.03$, the particles are scattered mainly because of the mirroring effect produced by the largest perturbed wavelengths. The differences between SoF and CoF solutions can be used to qualitatively estimate the impact of TTD over the CR transport: in a CoF geometry, the fraction of fast-magnetosonic waves is larger and the TTD contributes more to the scattering of particles. At higher rigidities, i.e. those interacting with turbulent perturbations in the inertial range, we can advance two arguments. First, considering only SoF solutions we see that the parallel mfp is reduced by a factor of $\sim$ 10 (with respect to the low rigidity regime); this shows that gyro-resonance by Alfv\'en waves can contribute to the pitch-angle scattering substantially. Second, if the SoF-to-CoF mfp ratio is reduced by a factor of $\sim$ 10, then the gyro-resonance process by Alfv\'en and fast waves contributes substantially. At higher Alfv\'enic Mach numbers the differences between the two forcing solutions vanish (the ratio $\lambda_{\parallel}(\chi=1)/\lambda_{\parallel}(\chi=0) \rightarrow 1$ whatever the rigidity regime). It is difficult at this stage to draw conclusion from this effect. It could be associated with a stronger mode conversion between Alfv\'en and fast-magnetosonic waves at scales smaller than the injection scales.\\ 
However, the above analysis is still limited and a quantitative analysis would require to isolate the effects of the different modes that compose the turbulent spectrum in the different forcing limits. One possibility would be to use the mode decomposition procedure proposed by \cite{2003MNRAS.345..325C} and improved by \cite{2010ApJ...720..742K}, which should be valid in the limit $M_a < 1$ (see also the cases discussed in \cite{2012ApJ...750..150W}).\\

\noindent{\it Perpendicular mean free paths:} Considering now figure \ref{F:Fig4}, we can first compare our results with the solutions obtained by \citet{2008ApJ...673..942Y}. In the SoF case, $\lambda_{\parallel}/L_{inj}$ varies in the range $1-10^3$ (see the upper part of figure \ref{F:Fig3}). At $\rho=0.01$ we have $\lambda_{\parallel}/L_{inj} > 100$ and $\lambda_{\perp} \propto M_a^{2.94\pm0.62}$. So the result is not compatible with a $M_a^4$ scaling, but is still close to it. At such low rigidities and because $\lambda_{\parallel}/L_{inj} \gg 1$, particles are expected to move along a field line over several coherence lengths with a curvilinear abscissa $s \sim v t$ and their diffusion perpendicular mean magnetic field is determined by the divergence of magnetic field lines.  Again, because wave-particle interactions occur in the turbulence dissipation range at these rigidities, it is not possible to fully trust that the derived parallel mfp is the one to be expected if the turbulence were fully developed at these scales. Including all the missing modes would likely produce a smaller parallel mfp. In turn, we cannot discard the possibility that particle scattering could modify the result obtained for $\lambda_{\perp}$ substantially even if we expect the effect to be small at low $M_a$. We do not find any clear trend of super-diffusive CR transport which could produce a larger final perpendicular mfps with respect to the diffusive regime. Super-diffusion is produced by the strong-Alfv\'enic turbulence \citep{2014ApJ...784...38L} in an interval which widens as $M_a$ increases whereas perturbed scales associated with weak-Alfv\'enic turbulence which produces magnetic field line diffusion are restricted to scales $\sim L_{inj}$ only. We think that we need a higher grid resolution at low $M_a$ in order to include scales where GS spectrum develops in order to capture a possible effect of super-diffusion.
At $\rho=0.097$, we find $\lambda_{\perp} \propto M_a^{2.52 \pm 0.26}$, so we have a slower Alfv\'enic Mach dependence at this rigidity. Again, the result is not compatible with a $M_a^4$ dependence. At high rigidities there is a competing effect between magnetic field line wandering and particle magnetic scattering. We have $\lambda_{\parallel}/L_{inj} \sim 1$, and not $>1$, so wave-particle interactions have stronger effect over the perpendicular transport. This may explains why our solutions are not completely compatible with $\lambda_{\perp} \propto M_a^4$. It is again difficult to isolate an effect of super-diffusion in that rigidity regime. At high $M_a$, we find a perpendicular mean free path about two times smaller than the prediction $\lambda_{\perp}/L_{inj} \sim 1/3 M_a^4$ in relative agreement with the limit found by \citet{2008ApJ...673..942Y} and in agreement with the results presented in XY13. In the CoF case, our solution is $\lambda_{\perp} \propto M_a^{1.42 \pm 0.37}$ (lower part of figure \ref{F:Fig4}). As $\lambda_{\parallel}$ is reduced with respect to the SoF case we also tested the relation $\lambda_{\perp} = \lambda_{\parallel} M_a^4$. We find that our result is also not completely compatible with the latter relation (although not so far because the extreme values of $\alpha$ are 1.79 and 1.85), respectively, but it is strongly incompatible with $\lambda_{\perp} \propto M_a^4$. Here again, the parallel mfps cover a range $(\sim 1-50)L_{inj}$. This may explain why we do not find a clear trend. The CoF cases are characterized by a stronger effect of particle scattering produced by fast-magnetosonic waves.  \\
In both forcing geometries our solutions bracket the QLT solution $\lambda_{\perp} \propto M_a^2$ (from above for the SoF solutions and from below for the CoF solutions). Our results are not completely compatible with such a scaling possibly because the last point at $M_a =0.89$ is a bit under (above) the extrapolation of $M_a^2$ in the CoF case (SoF case). At high $M_a$ indeed the use of QLT is questionable.\\

\noindent{\it Summary:} Our solutions are globally consistent with the results obtained by XY13 to the notable exception of $\lambda_{\parallel}$ at $M_a < 0.7$. This effect is possibly related to differences in forcing procedures and methods used to solve the MHD equations in the codes. If we accept that SoF (CoF) drives at large scales Alfv\'en (fast-magnetosonic) waves preferentially, our results can give some qualitative hints about the respective effect of gyro-resonance and TTD processes.  At low rigidity, i.e. for particles in resonance with modes in dissipation range of the turbulence, the TTD likely controls the parallel CR mfp. However at such rigidities the solutions have to be taken with care as resonant modes are missing at small scales owing to the finite resolution of the simulations. With respect to the low rigidity findings, at higher rigidities the mfp is considerably reduced in the SoF case, which can be due to the effect of gyro-resonance by the Alfv\'enic turbulence. Also, the ratio of the parallel SoF and CoF mfps decreases, which can be due to the effect of gyro-resonance of Alfv\'en and fast-magnetosonic waves over the CR transport. We find that $\lambda_{\parallel} \propto M_a^{-2}$ predicted by the QLT is compatible with CoF solutions, but not with with SoF solutions. Our results concerning $\lambda_{\perp}$ are less affirmative than the ones advanced by XY13 and do not show a clear trend at the moment. The results have not been found to be fully compatible with the theoretical predictions proposed in \cite{2008ApJ...673..942Y}. One explanation is that for each specific $\chi$ value the parallel mfp is not entirely in the regime $\lambda_{\parallel}/L_{inj} < 1$ or $\lambda_{\parallel}/L_{inj} > 1$. Finally, our solutions are between the two scalings: $\lambda_{\perp} \propto M_a^{4}$ and $\lambda_{\perp} \propto M_a^{2}$ (the latter expected from QLT). We did not find a clear trend in our results that could be associated with an effect of super-diffusion especially at low rigidities and low Alfv\'enic Mach numbers likely because GS turbulence is restricted to reduced scale lengths. More simulations with larger dynamics are needed in order to choose definitively from among the effects of particle scattering and any effect of magnetic field line transport and also to decide among the different theoretical models. In particular, results at $M_a \sim 0.1$ would greatly help to probe the QLT predictions further. Unfortunately, such low Afv\'enic Mach numbers require very long integration timescales to reach a diffusive plateau. We come back to this aspect in \S \ref{S:DISRL}.

\subsection{CR mean free paths: forcing effects}
\label{S:FOR}
\begin{figure}
\centering
 \includegraphics[width=6.cm, angle=-90]{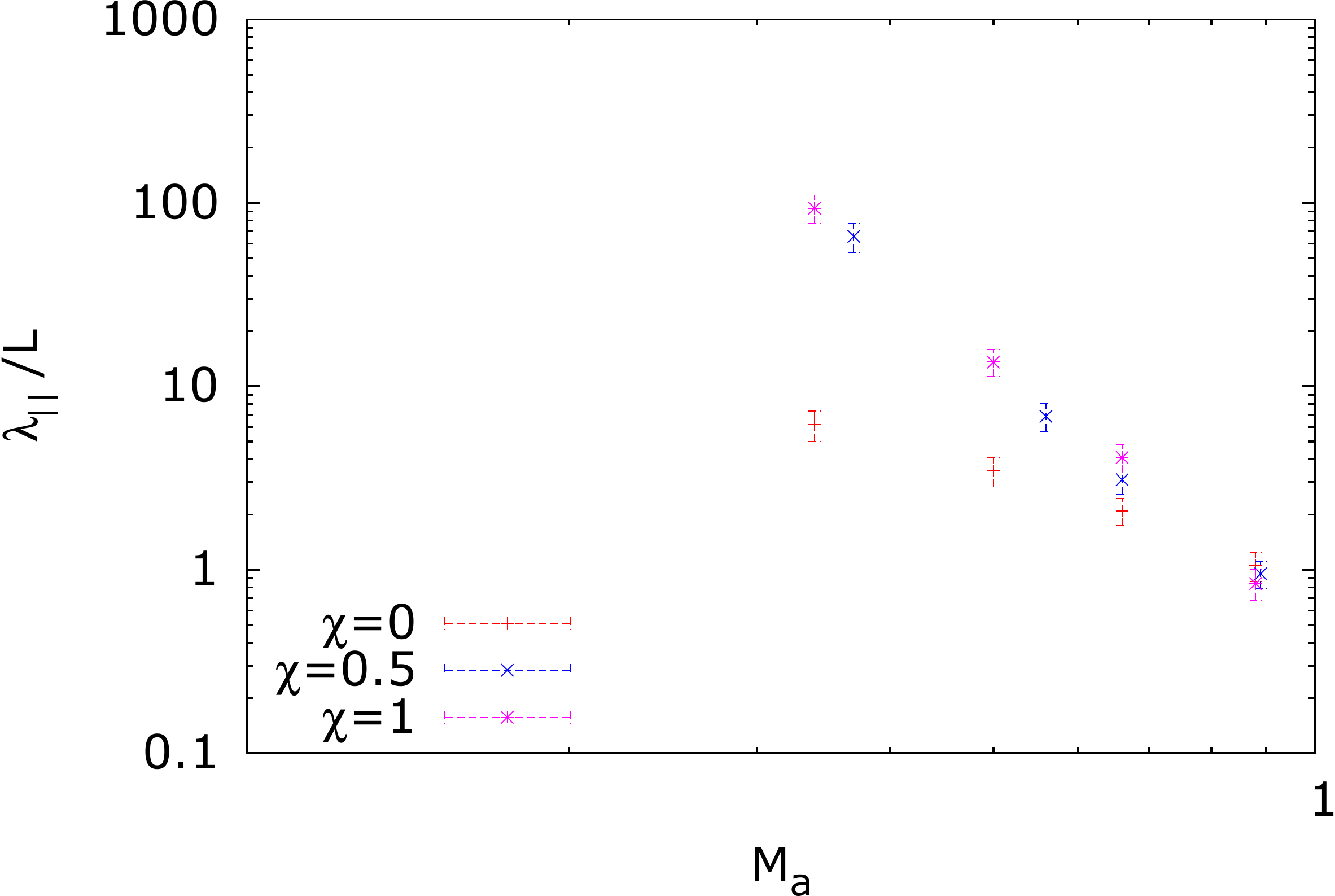}
 \includegraphics[width=6.cm,angle=-90]{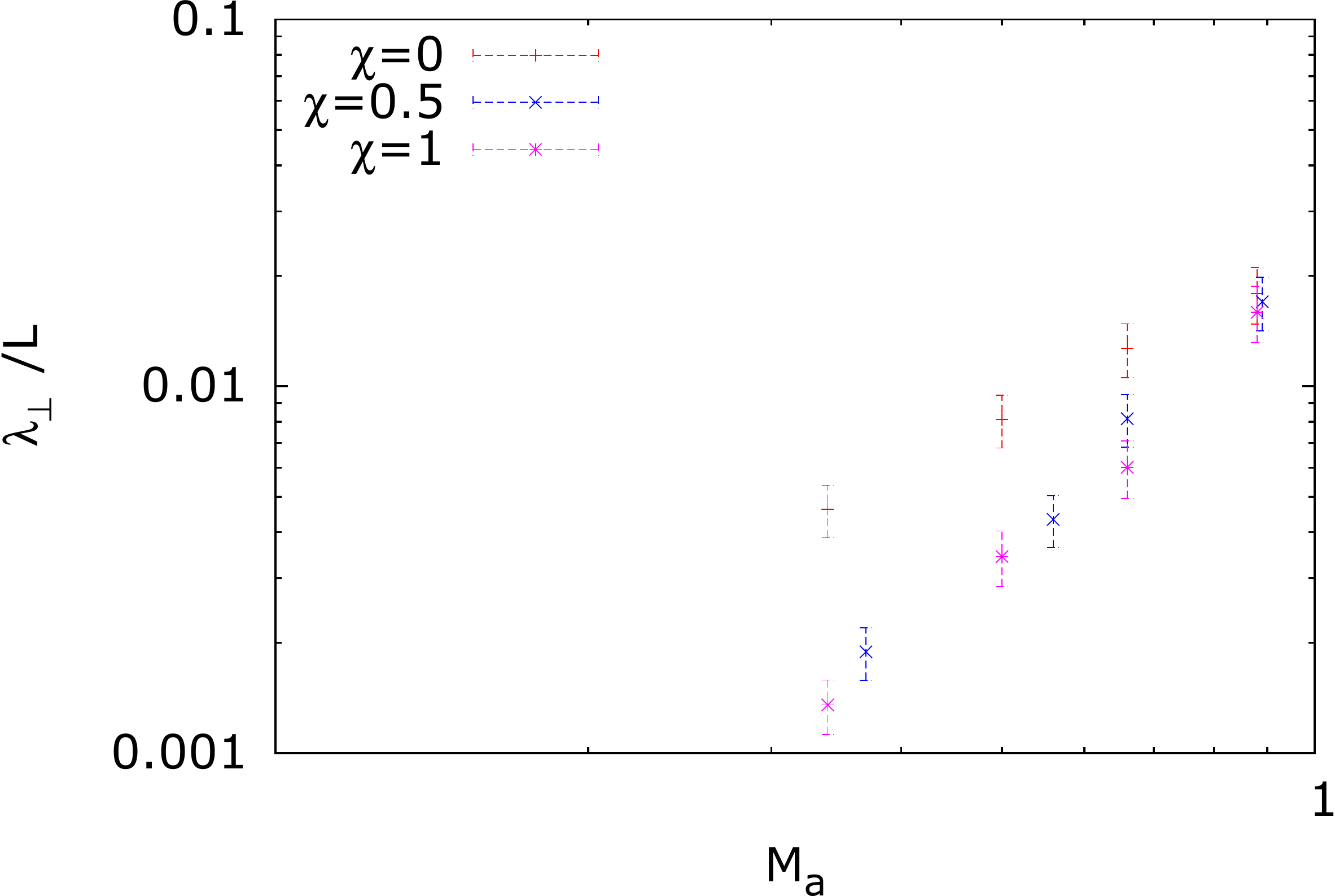}
 \caption{ [Upper figure]: Parallel mean free path dependence with $M_a$ at $\rho=0.097$ for $\chi=0,0.5,1$. [Lower figure]: Perpendicular mean free path dependence with $M_a$ at $\rho=0.097$ for $\chi=0,0.5,1$. } 
\label{F:Fig5}
\end{figure}

Figure \ref{F:Fig5} presents the dependence of parallel and perpendicular mfps on $M_a$ at $\rho= 0.097$ for $\chi=0,0.5$, and $1$.

It can be seen that in both cases (parallel and perpendicular mfps), especially for $M_a \le 0.5$, the MoF solutions at $\chi=0.5$ are closer to the SoF solutions. This is supported by the value of $r_c$ and the fraction of compressible modes (see \S \ref{S:Force}): in the MoF case $r_c=1/3$ is closer to $r_c=0$ obtained for $\chi=1$ than $r_c=1$ obtained for $\chi=0$. This fits with the arguments advanced above that Alfv\'en waves are preferentially produced with respect to fast magnetosonic waves in the SoF and MoF geometries and that a tiny fraction of the Alfv\'en energy is transferred to fast waves at low turbulent levels \citep{2003MNRAS.345..325C}. At higher Alfv\'enic numbers the effect of the forcing geometry is less stringent possibly because of a more efficient mode conversion. There, we find that parallel and perpendicular mfps are independent of $\chi$.

\subsection{CR mean free paths: Rigidity dependences}
\label{S:CRRHO}
\subsubsection{Resolution tests}
\label{S:RES}
\begin{figure}
\centering
 \includegraphics[width=6.cm, angle=-90]{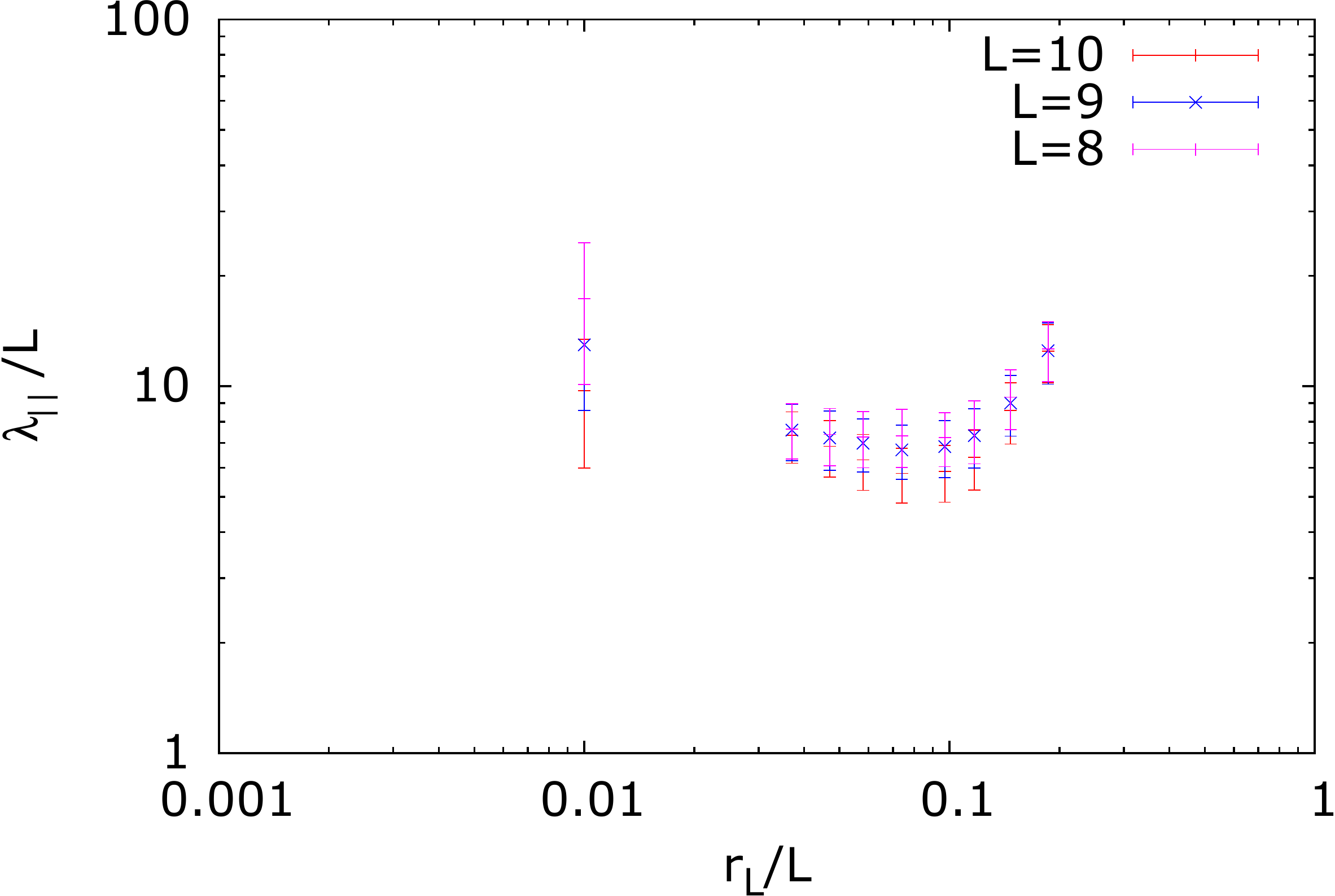}
 \includegraphics[width=6.cm, angle=-90]{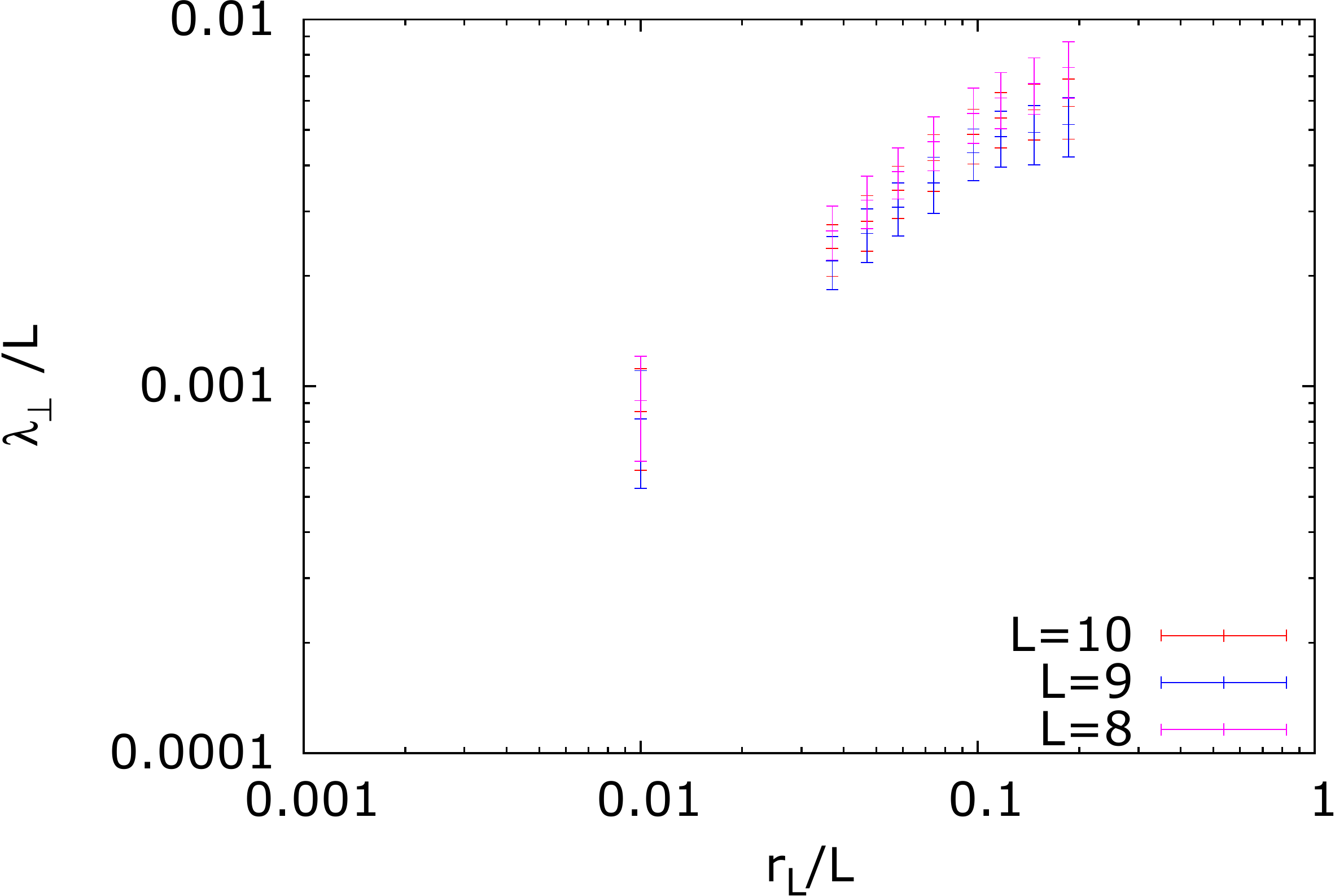}
 \caption{[Upper figure]: Parallel mfps at three different resolution for jobs 8J02c0.5, 9J06c0.5, and 10J01c0.5.   [Lower figure]: The same but for perpendicular mfps.} 
\label{F:Fig6}
\end{figure}

Before discussing any specific rigidity dependence we first test our calculation procedure for $\rho > 0.037$ at different grid resolution. Figure \ref{F:Fig6} presents the parallel and perpendicular mfp with respect to $\rho$ at X= 8, 9, and 10. All the simulations were performed at $\chi=0.5$ and jobs 8J02c0.5, 9J06c0.5, and 10J01c0.5 were used.\\
It can be seen that the results are compatible each other whatever the value of X. We also see a trend with a decreasing parallel mfp as the resolution increases, especially at rigidities $\rho < 0.1$. An effect that is expected since as X increases the inertial turbulence zone is shifted towards smaller scales.

\subsubsection{Parallel mean free paths}
\begin{figure}
\centering
 \includegraphics[width=6cm,angle=-90]{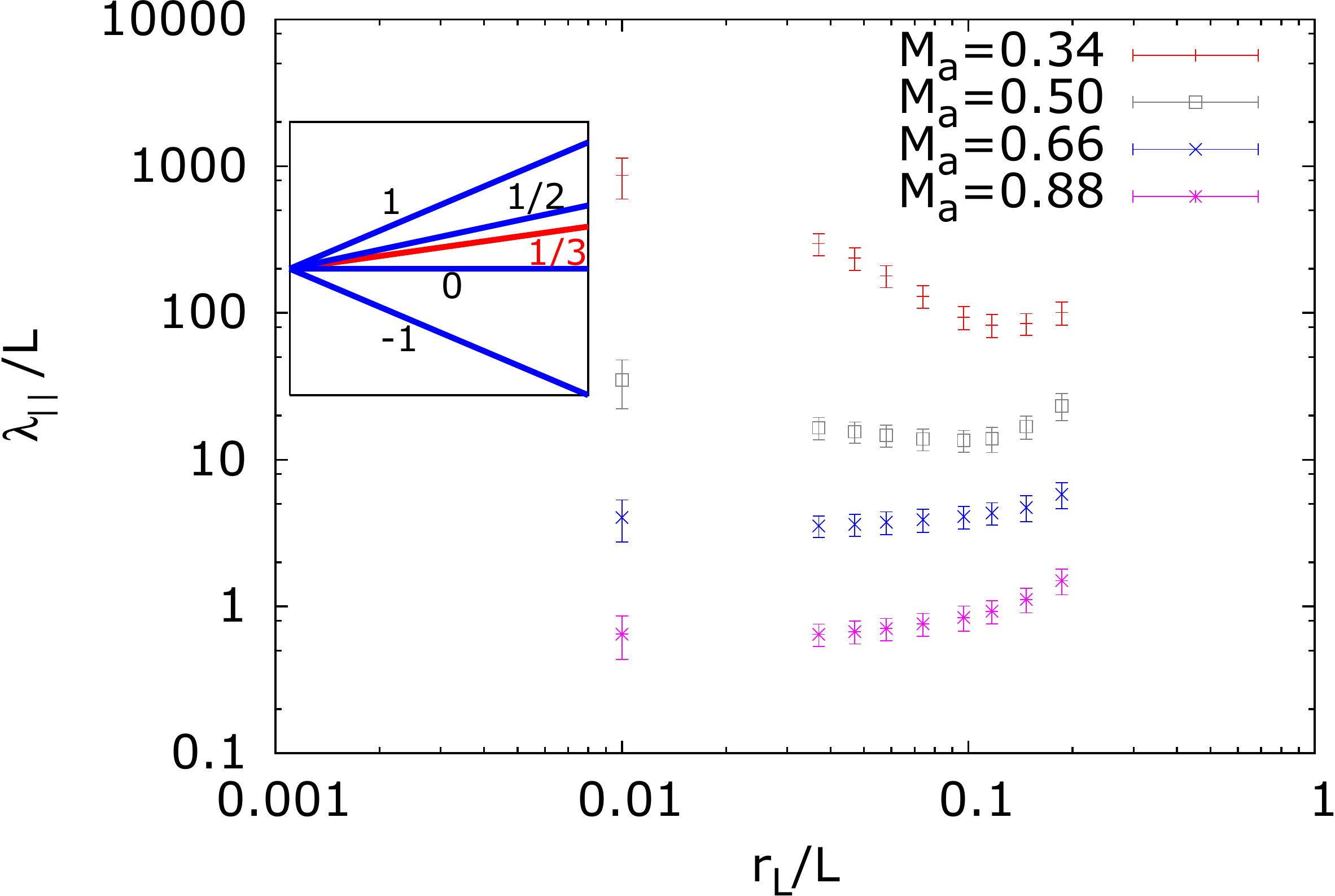}
 \includegraphics[width=6cm,angle=-90]{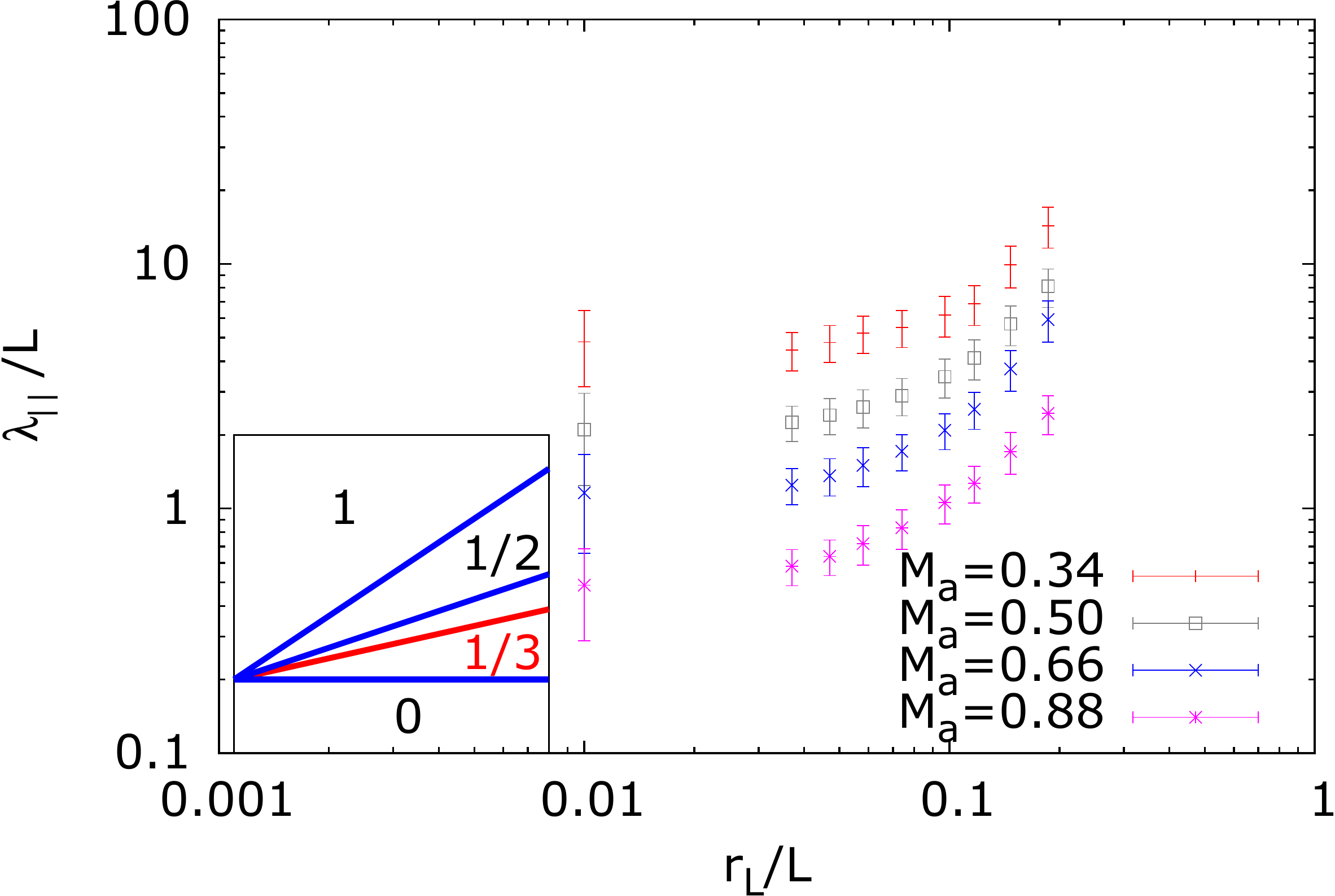}
 \caption{[Upper figure] Parallel mfps versus particle normalized Larmor radii $\rho$ at different $M_a$ in the SoF case ($\chi =1$). [Insets] Slopes obtained for different indices. [Lower figure] Parallel mfp versus particle normalized Larmor radius $\rho$ at different $M_a$ in the CoF case ($\chi =0$).} 
\label{F:Fig7}
\end{figure}

Figure \ref{F:Fig7} presents CR parallel mfps dependences with respect to the particle normalized Larmor radius $\rho=r_L/L$. \\

In table \ref{T:mfPPAR} we present the fit indices and the errors corresponding to linear fits of the parallel mfp for $0.01 <\rho < 0.1$ and $0.03 <\rho < 0.1$ in both SoF and CoF cases respectively.\\
\begin{table}[!ht]
\caption{Parallel mean free path indices and errors for SoF and CoF geometries.}             
\label{T:mfPPAR}
\centering                          
\begin{tabular}{c c c c}        
\hline\hline                 
Job no. & $M_a$ & $\alpha(0.03-0.1) \pm \Delta \alpha$ & $\alpha(0.01-0.1) \pm \Delta \alpha$  \\    
\hline                        
8J02c1.0 & 1.00 & $0.53 \pm 0.35$ & $0.34 \pm 0.16$    \\
9J09c1.0 & 0.34 & $-1.19 \pm 0.36$  & $-0.98 \pm 0.15$    \\
9J10c1.0 & 0.50 & $-0.21 \pm 0.35$  & $-0.36 \pm 0.20$    \\
9J11c1.0 & 0.67 & $0.27 \pm 0.47$  & $0.02 \pm 0.22$    \\
9J12c1.0 & 0.89 & $0.27 \pm 0.39$  & $0.13 \pm 0.23$    \\
10J02c1.0 & 0.37 &$-0.87 \pm0.35$ &$-0.64 \pm 0.13$ \\
\hline                        
9J01c0.0 & 0.35 & $0.34 \pm 0.39$  & $0.16 \pm 0.20$    \\
9J02c0.0 & 0.50 & $0.44 \pm 0.37$  & $0.29 \pm 0.21$    \\
9J03c0.0 & 0.66 & $0.54 \pm 0.35$  & $0.37 \pm 0.19$    \\
9J04c0.0 & 0.89 & $0.62 \pm 0.37$  & $0.44 \pm 0.18$    \\
\hline                                   

\end{tabular}
\end{table}

\subsubsection{Perpendicular mean free paths}
\begin{figure}
\centering
 \includegraphics[width=6cm,angle=-90]{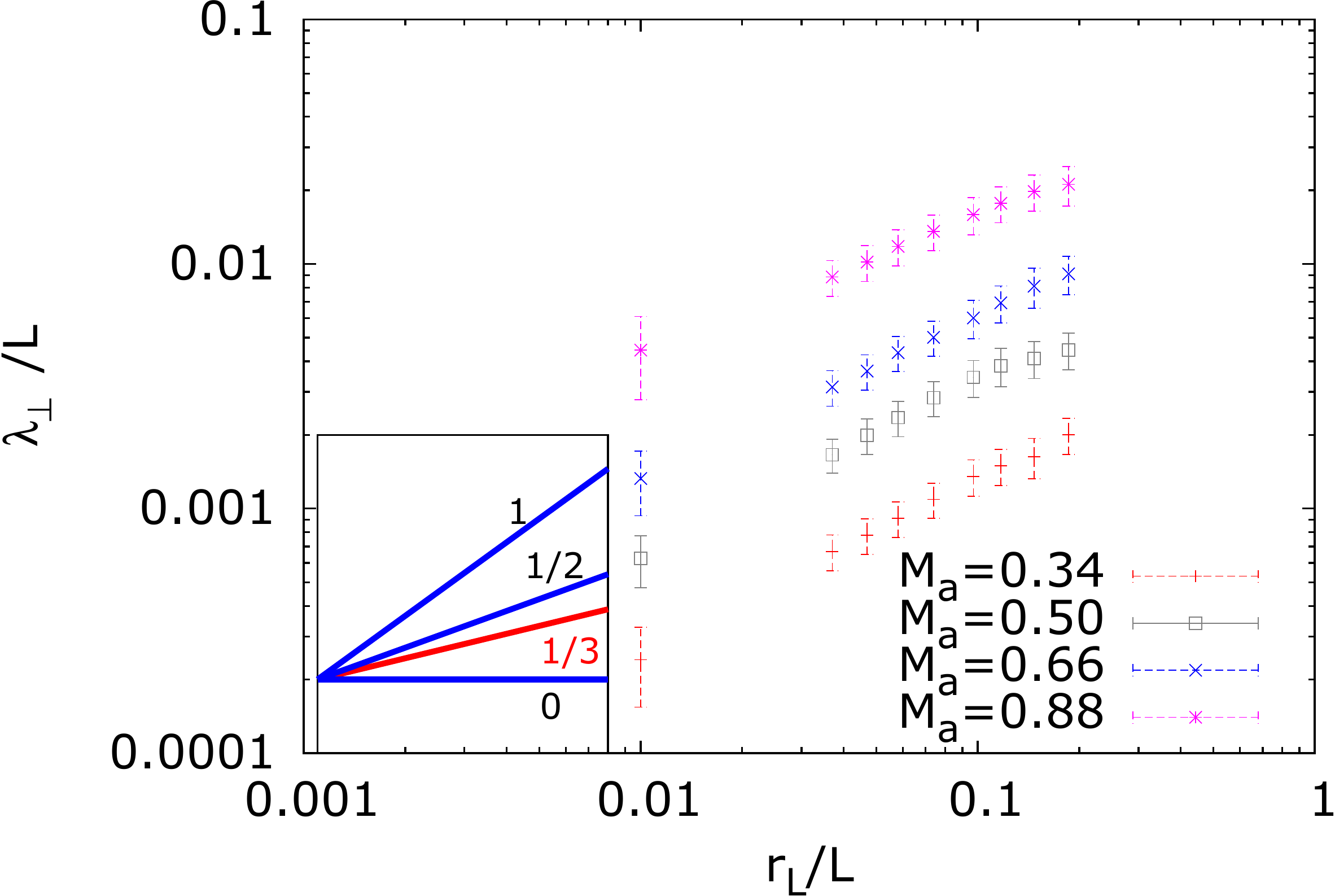}
 \includegraphics[width=6cm,angle=-90]{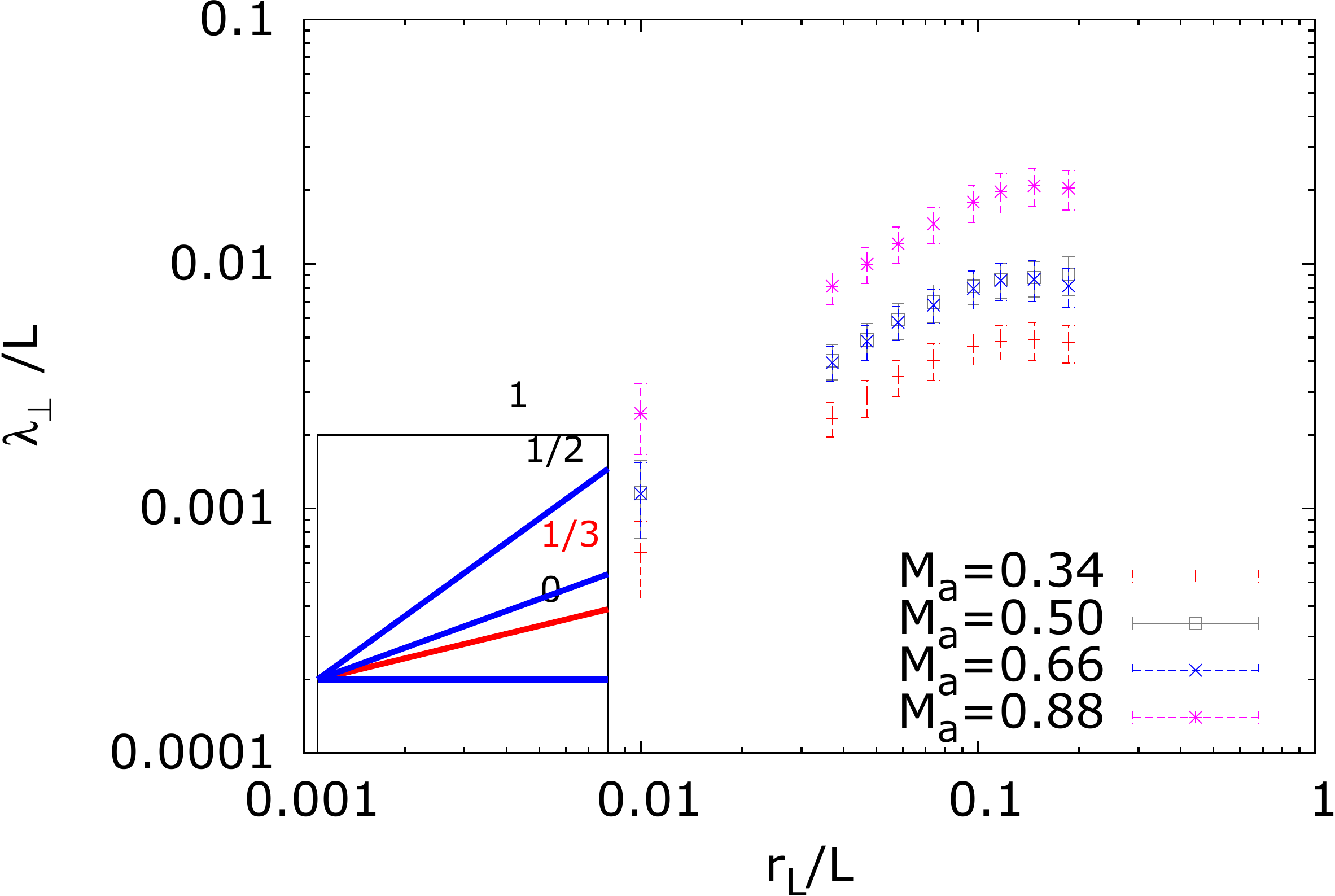}
 \caption{[Upper figure] Perpendicular mfp versus particle normalized Larmor radius $\rho$ at different $M_a$ values in the SoF case ($\chi =1$).
[Lower figure] Perpendicular mfp versus particle normalized Larmor radius $\rho$ at different $M_a$ values in the CoF ($\chi =0$).}
\label{F:Fig8}
\end{figure}

Figure \ref{F:Fig8} presents the CR perpendicular mfp with respect to the particle normalized Larmor radius $\rho=r_L/L$.\\

In table \ref{T:mfPPER} we present the fit indices and the errors corresponding to linear fits of the perpendicular mfp for $0.01 <\rho < 0.1$ and $0.03 <\rho < 0.1$  in the SoF and CoF cases respectively.\\
\begin{table}[!ht]
\caption{Perpendicular mean free path indices and errors for both forcing geometries. }             
\label{T:mfPPER}
\centering                          
\begin{tabular}{c c c c}        
\hline\hline                 
Job no. & $M_a$ & $\alpha(0.03-0.1) \pm \Delta \alpha$ & $\alpha(0.01-0.1) \pm \Delta \alpha$  \\  
\hline                        
8J02c1.0 & 1.00 & $0.56 \pm 0.36$ & $0.59 \pm 0.19$     \\
9J09c1.0 & 0.35 & $0.73 \pm 0.35$  & $0.78 \pm 0.24$    \\
9J10c1.0 & 0.50 & $0.76 \pm 0.35$  & $0.76 \pm 0.18$    \\
9J11c1.0 & 0.67 & $0.67 \pm 0.36$  & $0.68 \pm 0.21$    \\
9J12c1.0 & 0.89 & $0.61 \pm 0.36$  & $0.59 \pm 0.25$    \\
10J02c1.0 & 0.37 & $0.79 \pm 0.34$ & $ 0.78 \pm 0.23$\\
\hline                        
9J01c0.0 & 0.34 & $0.70 \pm 0.34$  & $0.86 \pm 0.19$    \\
9J02c0.0 & 0.50 & $0.73 \pm 0.34$  & $0.86 \pm 0.21$    \\
9J03c0.0 & 0.66 & $0.85 \pm 0.35$  & $0.94 \pm 0.18$    \\
9J04c0.0 & 0.89 & $0.82 \pm 0.35$  & $0.89 \pm 0.22$    \\
\hline                                   

\end{tabular}
\end{table}

\subsubsection{Discussion}
\label{S:DISRL}
Some obvious points can be highlighted. The SoF and CoF solutions for $\lambda_{\parallel}$ show very different rigidity variations. In the SoF case we see a clear trend that there is a spectral softening as $M_a$ decreases with some solutions showing inverted spectra (negative $\alpha$). In the CoF case no such effect is present and all the solutions have a spectral index largely independent of the Alfv\'enic Mach number. Solutions for $\lambda_{\perp}$ show similar spectra in the SoF and CoF cases, even if the SoF solutions have a stronger evolution with $M_a$ (see \S \ref{S:DISMA}).\\

\noindent{\it Comparison with previous works:} We first compare our results with the solutions obtained by BYL11. As reported above, BYL11 have derived their results in the incompressible MHD limit; that is in the limit $\beta_p = P_g/P_m \sim (M_a/M_s)^2 \rightarrow \infty$ \footnote{$\beta_p$ is the plasma parameter, $P_p$ and $P_m$ are the gas thermal pressure and the magnetic pressure respectively.}. As can be seen from table \ref{T:MHD}, most of our simulations are performed in an equipartition regime with $\beta_p \sim 1$. BYL11 also performed simulations at $M_a=1$ and $M_a=0.1$, but presented their results for $\lambda_{\parallel}$ at $M_a=1$ only (see \S \ref{S:Sum}). Hence, the more appropriate simulation to be compared with these results is the job 9J12c1.0. In that case, we find an index compatible with $\alpha=1/3$ and $\alpha=0$ (thus retaining or not the point at $\rho=0.01$) hence compatible with the BYL11 results (see table \ref{T:mfPPAR}). It should be noted here that the results at rigidities $\rho < 0.02$ presented in BYL11 correspond to particles propagating in the dissipation regime of the MHD turbulence even if the resolution of the MHD simulations performed by the authors is higher than ours ($768^3$ instead of $512^3$). Again, we think these results should be considered with some caution especially for $\lambda_{\parallel}$. For $\lambda_{\perp}$, $\lambda_{\perp} \propto \rho^{1/3}$ at $M_a=0.89$ can provide a reasonable fit of the points with $\rho < 0.1$ in the SoF case but still removing the point at $\rho = 0.01$. Hence it is also compatible with BYL11 results.  In order to push ahead the comparison with BYL11, we also performed a series of simulations at level 8 at high $\beta_p \sim 100$ (job 8J02c0.1) and $M_a=1$. Here the index is found to be compatible (although a bit larger) with $\alpha=1/3$, but not with $\alpha=0$. \\

\noindent {\it Theoretical frameworks:} QLT predicts $\lambda_{\parallel} \propto \rho^{2-s}$, where $s$ is the 1D index of the magnetic turbulent spectrum. In the case of a Kraichnan or a Kolmogorov spectrum with $s=3/2$ and $s=5/3$, QLT predicts $\alpha=1/2$ and $1/3$, respectively. However, such indices are not restricted to QLT solutions. For instance, solutions of CR transport in isotropic turbulence with a Kolmogorov spectrum have an index $\alpha=1/3$ irrespective of the turbulence level \citep{2002PhRvD..65b3002C}, a result that cannot be anticipated using the predictions of the QLT. One important question is wether this scaling is also recovered in a Goldreich-Sridhar (GS) type turbulence \citep{2006A&A...453..193M}. Indeed the GS spectrum follows a Kolmogorov scaling in the perpendicular wave number space. \cite{2015JGRA..120.4095H} performed simulations using different synthetic turbulence models and in particular the GS model. Their simulations have been performed in the strong turbulence limit with $\eta_0=\delta B/B_0 =1$. The authors found $\alpha \simeq 1/3-1/2$ (the fit over one decade in rigidity does not allow a value between the two scalings) and also found that GS turbulence is almost as efficient as the isotropic Kolmogorov turbulence to scatter CRs. \cite{Yan02} and \cite{2000PhRvL..85.4656C} found several orders of magnitude of differences between the CR mfps produced by isotropic and GS models for $M_a < 1$ \footnote{Aalthough the mfps obtained by \cite{Yan02} are 4 orders of magnitude above the ones obtained by \cite{2000PhRvL..85.4656C}.}. \cite{2008ApJ...673..942Y} instead reported on a dominance of fast-magnetosonic modes which follow an isotropic Kraichnan turbulent spectrum. Hence, if this result is correct, one can expect to find $\alpha=1/2$. We note that this result has been derived using a non-linear correction to the particle trajectory and hence it supersedes QLT solutions (see \cite{2008ApJ...673..942Y} for details). Finally, \cite{2015JGRA..120.4095H} obtained $\lambda_{\perp} \propto \rho^{1/2}$ in the case of GS spectrum at $\delta B = B_0$ (even if the fit is restricted to almost one decade in rigidity). At lower $M_a$, only if $\lambda_{\parallel} < L_{inj}$ \cite{2008ApJ...673..942Y} find a perpendicular mfp varying with the particle rigidity as $\lambda_{\parallel} M_a^4 \propto \rho^{1/2}$. \\

\noindent{\it Parallel mean free paths:} We now discuss each specific result in details. In all cases, the indices for the parallel mfp are found in the range $(-1.5,1)$. In particular, spectra harder than $\lambda_{\parallel} \propto \rho$ are rejected. \\
The lower part of figure \ref{F:Fig6} presents the parallel mfp in the CoF case. In that case, we never find indices compatible with the Bohm scaling ($\alpha = 1$) but the index at $M_a=0.89$ which is marginally compatible. The mfp index in the regime $\rho < 0.1$ shows some hardening from low to high $M_a$, except for in all cases indices are compatible either with 1/3 or 1/2 (see table \ref{T:mfPPAR}). This hardening trend is reinforced if we include the point at $\rho=0.01$ in the fit: the points at $\rho=0.01$ produce a bias towards smaller values of the indices. However, it is expected that if located in the inertial range of the turbulence, these points should be shifted towards smaller mfps. Overcoming this effect would have required simulations at levels larger than $X=10$, which is beyond the computing resources available for this work. \\
If we now turn to the SoF case (upper part of figure \ref{F:Fig7}) at low $M_a$, the results are clearly incompatible with $\alpha=1/2$ or even $1/3$ (see table \ref{T:mfPPAR}), and either when including or removing the points at $\rho=0.01$. For $M_a \le 0.5$ negative indices are obtained. This result is difficult to anticipate with the actual theoretical transport models. One possibility would be to invoke intermittency effects known to reduce the particle mean free path with respect to homogeneous turbulence \citep{2014ApJ...781...93A}. However, this possibility is questionable since we did not detect any subdiffusive paths at low rigidities and low $M_a$. Inverted indices can be associated with a propagation of CRs in a damped turbulence \citep{2008ApJ...673..942Y, 2010ApJ...725.2110S}. Above $M_a =0.5$ our results are compatible with $\alpha=1/3$ which is the expected result in Kolmogorov isotropic turbulence \citep{2002PhRvD..65b3002C}. \\

\begin{figure}
\centering
 \includegraphics[width=6cm, angle=-90]{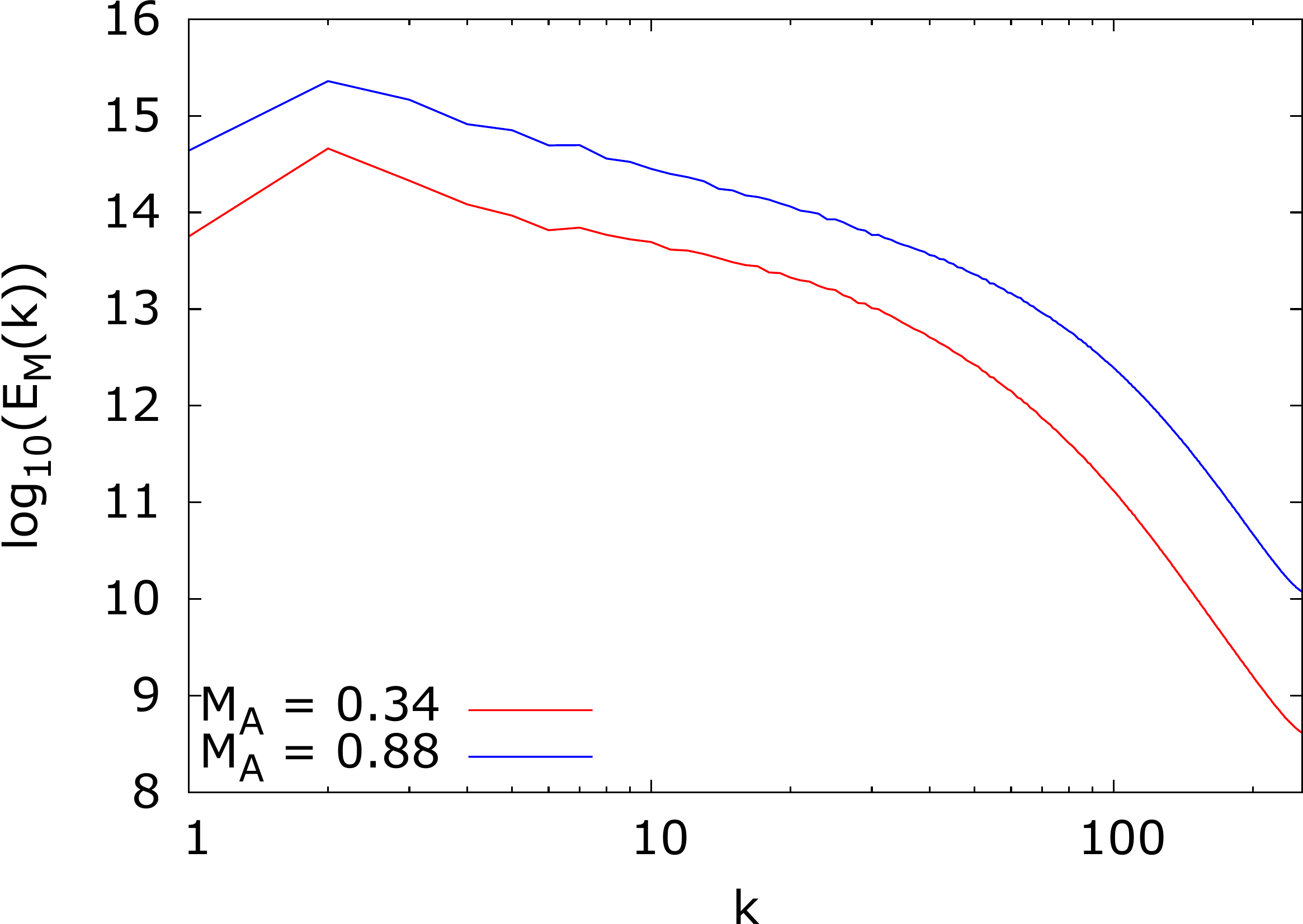}
 \caption{Magnetic energy density power spectra for $M_a=0.88$ (blue continuous line) and $M_a=0.34$ (red continuous line) in the CoF case.} 
\label{F:Fig9}
\end{figure}

From the above discussion we find that $\lambda_{\parallel} \propto \rho^{1/2}$ is clearly compatible with the simulations in the CoF case. Hence we find that in the CoF limit our results are consistent with a CR transport controlled by fast-magnetosonic waves as interpreted in the paradigm presented by \cite{2008ApJ...673..942Y}. This argument is similar to the conclusion reached from the Alfv\'enic Mach number analysis in \S \ref{S:DISMA}. However, our results are also consistent with $\alpha=1/3$ and a definite answer will require more intensive simulations with extended dynamical range in order to select bewteen the two models. In figure \ref{F:Fig9}, we plot the magnetic energy density power spectrum at two Alfv\'enic Mach numbers $M_a=0.34$ and $M_a=0.89$. We see that the power spectrum index is almost the same ($s \sim 1.5-1.6$), which can explain why we find a rigidity index $\alpha=1/2-1/3$ largely independent of $M_a$. \\

\begin{figure}
\centering
 \includegraphics[width=6cm, angle=-90]{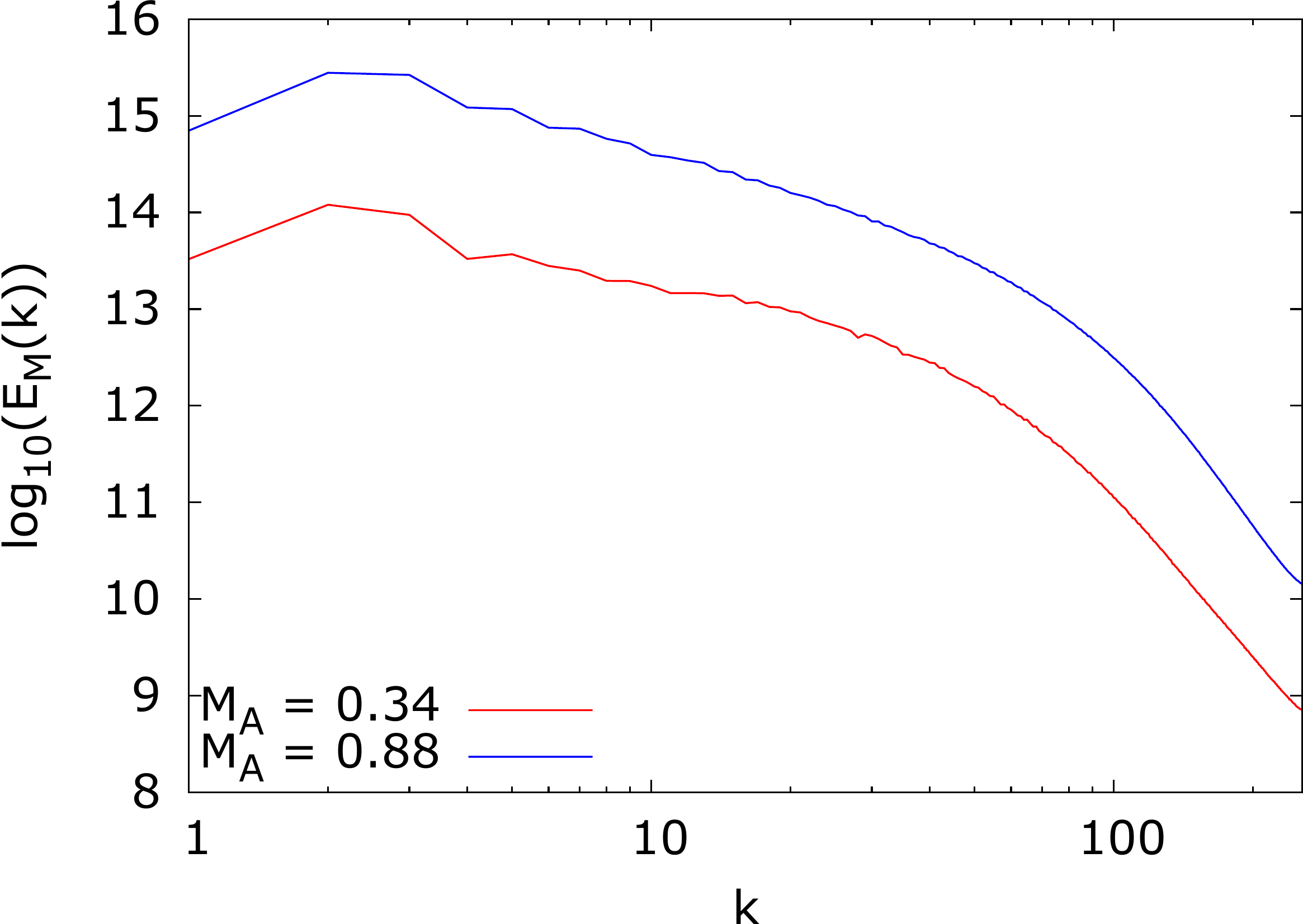}
 \caption{Magnetic energy density power spectra for $M_a=0.88$ (blue continuous line) and $M_a=0.34$ (red continuous line) in the SoF case.} 
\label{F:Fig10}
\end{figure}

The simulations obtained in the SoF case show very different trends. One way to get inverted spectra would be to have $s \ge 2$, hence the results could still be interpreted in the QLT framework. We plot in figure \ref{F:Fig10} the magnetic energy density power spectrum at $M_a=0.34$ and $M_a=0.89$. We checked that the stationarity was reached. If the spectrum at $M_a=0.89$ has an index $s \sim 1.5$ the spectrum at $M_a=0.34$ is very different: it is harder at large scales with $s \sim 1$ and softer above $kL \sim 18$ to $s \sim 2$. We note that scales $kL \sim 18$ correspond to rigidities $\rho \sim 0.05$ in figure \ref{F:Fig2} and this softening effect can explain the flat or negative values of $\alpha$. However, we cannot exclude with our simulations at level 9 that this spectral effect is of numerical origin; because of the finite resolution our simulations may have missed some resonant modes. In order to test this possibility we performed one simulation at $M_a=0.34$, but at level 10 (see table \ref{T:mfPPAR}). At X=10 we also find an inverted spectrum for the parallel mfp with an index compatible with the solution obtained at level 9, but smaller in absolute value (the effect is more pronounced if we include the point at $\rho=0.01$). This can be expected because low rigidity points with $\rho=0.01, 0.037$ and $0.047$ have lower parallel mfp at X=10. A higher resolution shifts the dissipation scale towards smaller scales and produces more fluctuations in resonance with low rigidity particles (see \S \ref{S:RES}). So, we conclude that inverted spectra probably do not result from a numerical artifact. It would be interesting to explore the $(M_a, M_s)$ parameter space to test if this kind of spectrum is a general trend. Finally, this effect can possibly be related with the behavior of weak sub-Alfv\'enic turbulence (see \S \ref{S:CRMA}). Hence, if $M_a=0.3$, weak sub-Alfv\'enic turbulence is present over one magnitude in $k$ beyond the injection scale. Because the energy cascades in the perpendicular direction, in that case, a low CR scattering efficiency and large parallel mfps are expected. \\

When considering parallel mfps above $\rho=0.1$, in all cases the spectrum hardens but our simulations do not have enough dynamics to probe the high-rigidity regime where $\lambda_{\parallel} \propto \rho^2$ is expected. This rigidity dependence is expected in the configuration of CR propagation with Larmor radii larger than the turbulence coherence length (note that $L_{inj} \sim 0.2$) (see \cite{2004APh....21..609D, 2013MNRAS.430.1280P}). We also note that a transition between small to large scale turbulence at $\rho=0.1$ gives a transition Larmor radius $r_L \sim 0.5 L_{inj}$ as $L_{inj} = 0.2 L$. \\

\noindent {\it Perpendicular mean free paths:} Considering now perpendicular mfps (see figures \ref{F:Fig7}), the CoF solutions are not compatible with $\alpha=1/3$, but are compatible with $\alpha=1/2$. The SoF solutions show spectral indices all compatible with $\alpha=1/2$ with or without the points at $\rho=0.01$. In fact, if we remove these points the results are also compatible with $\alpha=1/3$ (this may be less true at low $M_a$).  At low $M_a$, the index is also compatible with the solution obtained at level 9 (see table \ref{T:mfPPER}). At $M_a=0.66$ and $M_a=0.89$, we find that $s \sim 1.6$ fits the data well leading to $\alpha=1/3$, but the spectral dynamics is too restricted again to separate solutions with $\alpha=1/2$ or $1/3$. At $M_a \sim 0.89$, we also find that the mfps in the CoF and SoF cases are similar. The interpretation of this is difficult and require requires distinguishing the impact of each type of mode (see \S \ref{S:DISMA}). It is likely that $\lambda_{\perp}$ at high $M_a$, whatever the forcing, is controlled by the field line wandering process, which is an issue difficult to test using an analytical approach (see discussions in \citep{2002PhRvD..65b3002C, 2010ApJ...720L.127S}). We note that the wandering of the field lines is controlled by the shearing effect produced during the interaction of two counter-propagating Alfv\'en wave packets. The perpendicular mfp above $\rho=0.1$ is compatible with a flat dependence on $\rho$ (see \cite{2002PhRvD..65b3002C, 2015JGRA..120.4095H}).\\

\noindent{\it Summary:} We again find that CoF solutions are compatible with QLT predictions (at least at low $M_a$) but also with non-linear models of CR diffusion in fast-magnetosonic turbulence. However, the dynamics in rigidity scales is not sufficient to differentiate between solutions with $\alpha =1/3$ or $\alpha = 1/2$. SoF solutions, however do not fit in this scenario; they are incompatible with QLT solutions at low $M_a$ and show spectra with a milder or inverted rigidity dependence that are connected with particular magnetic field spectra possibly connected with the transition between weak and strong sub-Alfv\'enic turbulence regimes. Perpendicular mfps show similar rigidity dependences in both SoF and CoF cases. There the wandering of field lines is likely important even if not dominant.

\subsection{The $\lambda_{\perp}/\lambda_{\parallel}$ ratio}
\label{S:RAT}
\begin{figure}
\centering
 \includegraphics[width=6cm, angle=-90]{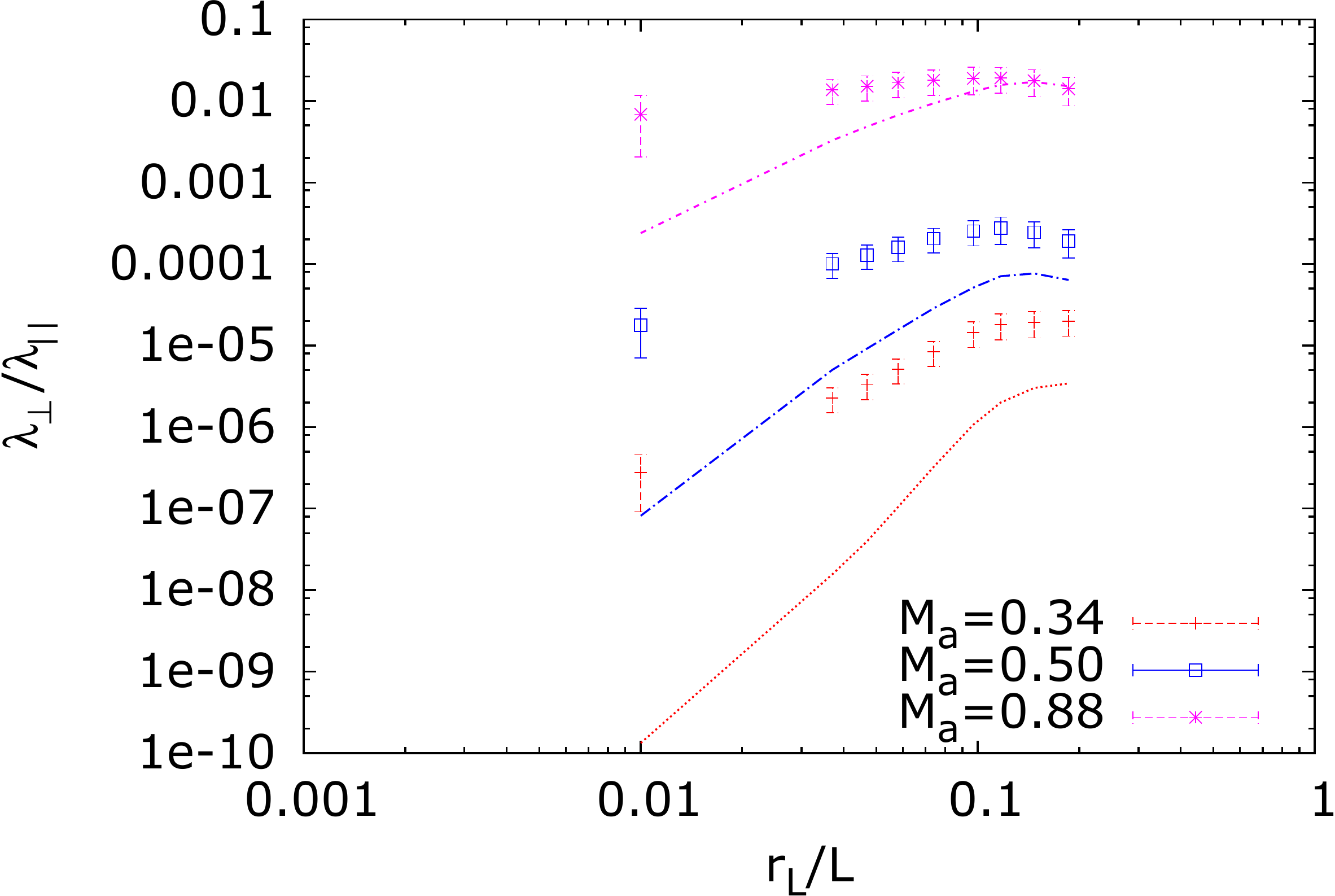}
 \includegraphics[width=6cm, angle=-90]{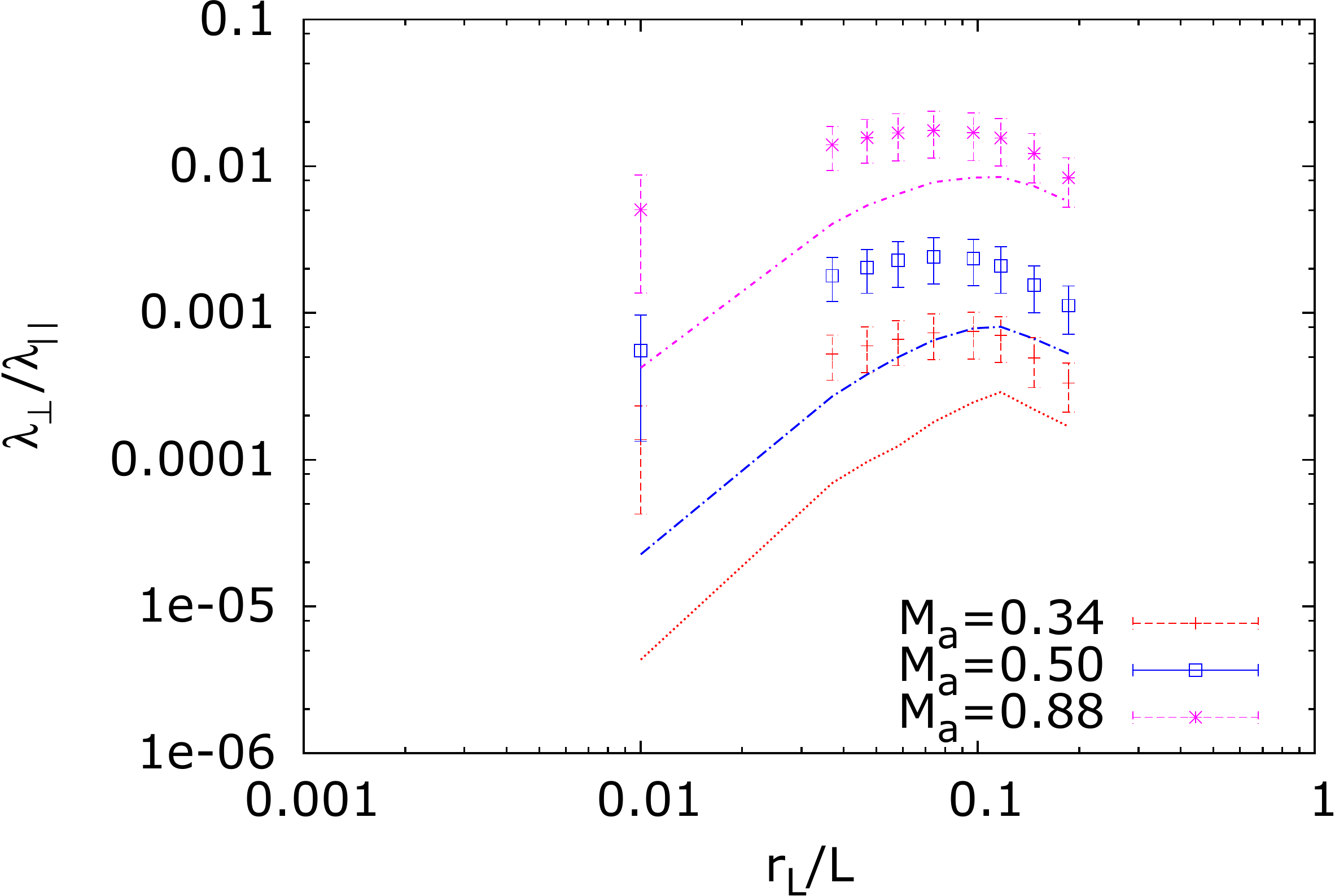}
 \caption{Ratios of $\lambda_{\perp}/\lambda_{\parallel}$ for different values of the Alfv\'enic Mach number. [Upper figure] SoF case. [Lower figure] CoF case. The continuous curves display the function  $1 / \left(1 + \left(\lambda_{\parallel} / r_L\right)^2\right)$.} 
\label{F:Fig11}
\end{figure}

Figure \ref{F:Fig11} presents the ratio $\lambda_{\perp}/\lambda_{\parallel}$ with respect to $\rho$ derived at $M_a=0.34, 0.66$ and $0.89$ in the SoF and CoF cases. This diagnostic is interesting for probing the importance of field line wandering over the CR transport (see \cite{2002PhRvD..65b3002C}). To that end it is first instructive to compare the latter ratio with the prediction from classical scattering theory \citep{1975Ap&SS..32...77F}: 
\beq
\label{Eq:CST}
{\lambda_{\perp} \over \lambda_{\parallel}} = {1 \over 1 + \left({ \lambda_{\parallel} \over r_L}\right)^2} \ .
\eeq
The solution given by Eq. \ref{Eq:CST} is displayed as the continuous line in the two panels in figure \ref{F:Fig11}. In both the SoF and CoF cases Eq. \ref{Eq:CST} is not compatible with our results (maybe with the exception of high-rigidity points where the contribution of pitch-angle scattering to $\lambda_{\perp}$ should be the strongest). The difference amounts to the contribution of the field line wandering process over the CR perpendicular transport. If we remove the point at $\rho=0.01$ our results are compatible with a flat ratio for $\rho < 0.1$ in the CoF case and in the SoF but only at $M_a =0.89$. If $\chi = 1$ and $M_a < 0.89$ a flat dependence is disfavored. Above $\rho =0.1$ the dynamics is reduced but the ratio is compatible with $\rho^{-2}$ especially in the CoF case. A flat ratio is expected if the field line wandering process takes over particle scattering.  It seems that a reduced efficiency of the parallel transport in sub-Alfv\'enic turbulence could provide an explanation of this spectral hardening at low CR rigidities. A more quantitative discussion on field line wandering requires a reconstruction of the magnetic field lines in different MHD realizations. This is the subject of a work in preparation. 

\section{Summary and conclusions}
\label{S:Conc}

In this work we have developed a series of kinetic-MHD simulations of CR transport in magnetized turbulence. We have upgraded the {\tt RAMSES} code to include a module that permits forcing the velocity space following different geometries and a particle-in-cell module which allows the reconstruction of the trajectory of charged particles in an electromagnetic field. We have investigated the Alfv\'enic Mach number and rigidity dependence of both parallel and perpendicular mfps by sampling a large number of particles and averaging over several statistically independent field realizations. \\

The main results of this work are the following:
\begin{itemize}

\item Forcing effects: The effects of the forcing geometry are the strongest at low Alfv\'enic Mach numbers where SoF produce parallel mfp that can be two orders of magnitude larger than CoF mfp solutions. A MoF produces results close to the SoF solutions. At high Alfv\'enic Mach numbers, the impact of the forcing geometry vanishes and all the results
converge to a similar solution. 

\item Alfv\'enic Mach number dependence: In the SoF case, we found faster $M_a$ dependences with respect to the results of XY13, but our results are compatible at $M_a > 0.7$. Even if we found a similar trend to XY13 for the dependence of $\lambda_{\perp}$ on $M_a$, we consider that these results have to be taken with care because these correspond to particle rigidities lying in the dissipation range of the turbulence. The missing resonant interactions are expected to reduce $\lambda_{\parallel}$ and possibly modify the result on $\lambda_{\perp}$. SoF results are not found to be compatible with QLT predictions. In the CoF case, we found a scaling compatible with $M_a^{-2}$ and so compatible with QLT predictions but for an Alfv\'enic Mach number regime beyond the standard validity domain of the QLT. SoF perpendicular mfps are not found to be compatible with the theoretical calculations proposed by YL08, even if our SoF solutions are close to a $M_a^4$ scaling. This can be explained because the ratio $\lambda_{\parallel}/L_{inj}$ is neither $\gg 1$ nor $\ll 1$ in our solutions. We did not find a clear signature of super-diffusion in the CR perpendicular mfp produced by GS-type turbulence possibly because of the limited resolution level. CoF perpendicular mfps have scalings relatively close to the QLT expectation owing to the wandering of magnetic field lines. 

\item Rigidity dependence: CoF gives results that are compatible with $\lambda_{\parallel} \propto \rho^{1/2}$, consistent with predictions from QLT and the diffusion produced by a wave spectrum following a Kraichnan law. This solution is expected if fast-magnetosonic waves dominate the CR scattering process and is valid even beyond QLT calculations. However, our results are also compatible with $\lambda_{\parallel} \propto \rho^{1/3}$. Simulations with larger dynamics are necessary in order to differentiate bewteen the two scalings. SoF show very different results, in particular at low Alfv\'enic Mach numbers where inverted or flat spectra are obtained. This result cannot be explained by QLT predictions. The effect may result from an inefficient CR scattering in weak sub-Alfv\'enic turbulence. Perpendicular mfps have rigidity dependences also compatible with $\rho^{1/2}$ for both CoF and SoF cases. CoF and SoF solutions are similar at high $M_a$. At rigidities $\rho > 0.1$, that is at rigidities larger than the turbulence injection scale, a hardening (a softening) of the parallel (perpendicular) mean path is obtained compatible with the propagation of particles in small-scale turbulence.

\item Ratio of perpendicular to parallel mfps:  A flat ratio characteristic of the effect of magnetic field line wandering at rigidities $\rho < 0.1$ is found in both CoF and SoF cases at high $M_a$. CoF solutions show the same trend whatever the level of turbulence whereas SoF solutions due to enhanced parallel mfp show a rigidity dependent ratio at low Alfv\'enic Mach number. Here weak scattering effects observed for $\lambda_{\parallel}$ seem to be as important as field line wandering to explain a smaller ratio at low rigidities.

\end{itemize}

Several issues emerge from this work. The solutions are found to be dependent on the adopted forcing geometry. It appears that CoF simulations are compatible with expectations from QLT or from some non-linear models, whereas SoF solutions are not. We have proposed some arguments to explain the parallel mfp in the SoF and CoF geometries. They are based on a preferential mode type production and the relative effect of TTD and gyro-resonance. However, this is speculation and it seems important at this stage to have a proper test of the effect of each type of MHD mode on the CR scattering process. This approach is important, for instance, to assert the inverted spectra obtained at low Alfv\'enic Mach numbers in the SoF case. To that end, it would be interesting to filter the different modes out of the magnetic field realizations and to repeat particle propagation runs. This aspect deserves a future investigation. This issue is also connected to the scattering efficiency in an Alfv\'enic turbulence which has been advanced to be weak because of the anisotropy but has never been tested by direct numerical simulations. It also seems important to extend the calculations of magnetic field transport in the low $M_a \le 0.1$ Alfv\'enic Mach number regime. It would be also interesting to perform high-resolution level simulations with X beyond 10 in order to be able to capture and disentangle the effect of strong and weak turbulence regimes over CR perpendicular mean free paths. These calculations however, require dedicated important computational resources that deserve future work. Extension of the present study to the trans- and super-Alfv\'enic ($M_a \ge 0.9$) cases would also be interesting in order to test the different models describing the propagation of particles in isotropic turbulence. Finally, it is also important to extend the simulations at low and high $\beta_p$ regimes as most of the jobs in this work were performed at $\beta_p \sim 1$. The magnetic field line transport process is important in order to understand the perpendicular mfp of the particles. Dedicated calculations of field line calculations in different forcing geometries are in progress and deserve a forthcoming work. \\
Finally, it should be emphasized that this study is restricted to the transport of particles in large-scale turbulence which is, for instance the case of high-energy CRs (TeV and beyond) in the ISM. The transport of low-energy particles in the ISM requires adding a component of self-generated waves (see the discussion in \S 6 of XY13). As the pressure imparted in the low-energy CR is as important as the gas and magnetic pressure, this would be possible using PIC-MHD simulations, but only if the source terms associated with the CR currents are correctly implemented in the MHD equations (see \cite{2014arXiv1412.1087B}). The corresponding upgrading of the PIC and MHD modules is inprogress.

%
  
%
  
\begin{acknowledgements}
We thank M. Lemoine, P. Hennebelle and H.Yan for fruitful discussions. We thank the anonymous referee for his/her references to the weak turbulence regime and super-diffusive effects. The simulations have been performed using the super calculator facilities available at the CINES (Centre d'Informatique de l'Enseignement Sup\'erieur in Montpellier) center. This work has benefit from the support of the ANR COSMIS project. 
\end{acknowledgements}

\bibliography{AM.bib} 
\bibliographystyle{aa}
\end{document}